\documentclass[fleqn,usenatbib]{mnras}

\usepackage{newtxtext,newtxmath}

\usepackage[T1]{fontenc}
\usepackage{ae,aecompl}

\usepackage{graphicx}	
\usepackage{amsmath}	

\usepackage{amssymb}	
\usepackage{subfigure}

\title[SRT observations of the Coma Cluster]{Sardinia Radio Telescope observations of the Coma Cluster}

\author[M. Murgia et al.]{
M. Murgia,$^{1}$\thanks{E-mail: matteo.murgia@inaf.it}
F. Govoni,$^{1}$
V. Vacca,$^{1}$
F. Loi,$^{1}$
L. Feretti,$^{2}$
G. Giovannini,$^{2,3}$
A. Melis,$^{1}$
\newauthor 
R. Concu,$^{1}$
E. Carretti,$^{2}$
S. Poppi,$^{1}$
G. Valente$^{1,4}$
G. Bernardi,$^{2,5,6}$ 
A. Bonafede,$^{3,2}$
\newauthor 
W. Boschin,$^{7,8,9}$
M. Brienza,$^{10,3}$
T.\ E. Clarke,$^{11}$
F. de Gasperin,$^{2}$
T. A. En{\ss}lin,$^{12}$
\newauthor 
C. Ferrari,$^{13}$
F. Gastaldello,$^{14}$
M. Girardi,$^{15,16}$
L. Gregorini,$^{2}$
M. Johnston-Hollitt,$^{17}$
\newauthor 
E. Orr{\`u},$^{18}$
P. Parma,$^{2}$
R. A. Perley,$^{19}$
G.B Taylor,$^{20}$
and P. Marchegiani$^{1}$
\\
\\
$^{1}$ INAF-Osservatorio Astronomico di Cagliari, Via della Scienza 5, 09047 Selargius, Italy\\
$^{2}$ INAF-Istituto di Radioastronomia, via P. Gobetti 101, 40129 Bologna, Italy\\
$^{3}$ Dipartimento di Fisica e Astronomia, Università di Bologna, via P. Gobetti 93/2, 40129, Bologna, Italy\\
$^{4}$ Agenzia Spaziale Italiana (ASI), Roma\\
$^{5}$ Department of Physics and Electronics, Rhodes University, PO Box 94, Grahamstown 6140, South Africa\\
$^{6}$ South African Radio Astronomy Observatory, Black River Park, 2 Fir Street, Observatory, Cape Town, 7925, South Africa\\
$^{7}$ Fundaci\'on G. Galilei - INAF TNG, Rambla J. A. Fern\'andez P\'erez 7, E-38712 Bre\~na Baja (La Palma), Spain\\ 
$^{8}$ Instituto de Astrof\'{\i}sica de Canarias, C/V\'{\i}a L\'actea s/n, E-38205 La Laguna (Tenerife), Spain\\
$^{9}$ Dep. de Astrof\'{\i}sica, Univ. de La Laguna, Av. del Astrof\'{\i}sico Francisco S\'anchez s/n, E-38205 La Laguna (Tenerife), Spain\\
$^{10}$ INAF – Osservatorio di Astrofisica e Scienza dello Spazio di Bologna, Via P. Gobetti 93/3, 40129 Bologna, Italy\\
$^{11}$ U.S.\ Naval Research Laboratory, Washington, District of Columbia 20375, USA\\ 
$^{12}$ Max Planck Institut f\"{u}r Astrophysik, Karl-Schwarzschild-Str.1, D-85740 Garching, Germany\\
$^{13}$ Laboratoire Lagrange, UCA, OCA, CNRS, Blvd de l'Observatoire, CS 34229, 06304 Nice cedex 4, France\\
$^{14}$ INAF - IASF Milano, Via Bassini 15, I-20133 Milano, Italy \\
$^{15}$ Dip. di Fisica, Universit\`a degli Studi di Trieste - Sezione di Astronomia, via Tiepolo 11, I-34143 Trieste, Italy \\
$^{16}$ INAF - Osservatorio Astronomico di Trieste, via Tiepolo 11, I-34143 Trieste, Italy\\
$^{17}$ International Centre for Radio Astronomy Research (ICRAR), Curtin University, Bentley, WA 6102, Australia\\
$^{18}$ ASTRON, the Netherlands Institute for Radio Astronomy, Postbus 2, 7990 AA, Dwingeloo, The Netherlands\\
$^{19}$ National Radio Astronomy Observatory, P.O. Box O, Socorro, NM 87801, USA\\
$^{20}$ Department of Physics and Astronomy, University of New Mexico, Albuquerque NM, 87131, USA\\
}
\date{Accepted XXX. Received YYY; in original form ZZZ}

\pubyear{2020}

\begin{document}
\label{firstpage}
\pagerange{\pageref{firstpage}--\pageref{lastpage}}
\maketitle

\begin{abstract}
We present deep total intensity and polarization observations of the Coma cluster at 1.4 and 6.6\,GHz performed with the Sardinia Radio Telescope. By combining the single-dish 1.4\,GHz data with archival Very Large Array observations we obtain new images of the central radio halo and of the peripheral radio relic where we properly recover the brightness from the large scale structures. At 6.6\,GHz we detect both the relic and the central part of the halo in total intensity and polarization. These are the highest frequency images available to date for these radio sources in this galaxy cluster. In the halo, we find a localized spot of polarized signal, with fractional polarization of about 45\%. The polarized emission possibly extends along the
north-east side of the diffuse emission. The relic is highly polarized, up to 55\%, as usually found for these sources. We confirm the halo spectrum is curved, in agreement with previous single-dish results. The spectral index is $\alpha=1.48\pm0.07$ at a reference frequency of 1\,GHz and varies from $\alpha\simeq 1.1$, at 0.1\,GHz, up to $\alpha\simeq 1.8$, at 10\,GHz. We compare the Coma radio halo surface brightness profile at 1.4\,GHz (central brightness and e-folding radius) with the same properties of the other halos, and we find that it has one of the lowest emissivities observed so far.  Reanalyzing the relic's spectrum in the light of the new data, we obtain a refined radio Mach number of $M=2.9\pm0.1$.
\end{abstract}




\section{Introduction}

Galaxy clusters are unique laboratories for the investigation
of turbulent fluid motions, large-scale magnetic fields, and relativistic
particles. Radio halos and relics, respectively located at the center and in the outskirts of galaxy clusters, 
provide direct evidence of the presence of relativistic particles and magnetic fields 
associated with the intra-cluster medium.
These diffuse sources, with no obvious optical counterparts, are characterized by a faint synchrotron emission, with steep radio spectra\footnote{$S(\nu)\propto \nu^{- \alpha}$, with
$\alpha$=spectral index} ($\alpha \gtrsim 1$), which extends over Mpc-scales\
\cite[e.g.][]{feretti12,brunetti14,vanweeren19}. 

Radio halos and relics are usually present in dynamically disturbed galaxy clusters, and their origin and evolution are likely to be associated with the merger history of the clusters. It is supposed that part of the gravitational energy dissipated by merger events is channelled into shocks and turbulence, which can provide the energy for acceleration of the relativistic electrons and protons as well as amplification of magnetic fields on large scales.

Central radio halos have been observed in a fraction of massive clusters and may probe turbulence dissipation. They are typically not polarized, 
and so far only a few examples (A2255, MACS\,J0717.5+3745, and A523) of large-scale filamentary polarized structures possibly associated with the radio halo emission have been detected \citep{govoni05,bonafede09a,girardi16,vacca23}. In the case of A2255, \citet{govoni06} suggest that these polarized filaments originate in the periphery of the cluster due to a peculiar steepening of the magnetic field power spectrum with radius. According to \citet{pizzo11}, however, these structures could alternatively be interpreted as foreground radio relics superimposed on the radio halo due to projection effects. A similar argument has been proposed by \citet{rajpurohit21} for the case of the cluster MACS J0717.5+3745, where 
the relic and a narrow-angle-tailed radio galaxy are interpreted as two different structures projected along the same line of sight.

In contrast to central radio halos, radio relics are highly polarized \citep[e.g.][]{clarke06,bonafede09b,vanweeren10,loi20}.
It is generally assumed that peripheral radio relics mark the location of shock fronts resulting from cluster merger events. The shock waves can amplify particle energies 
through the Fermi-I process, as well as amplify the magnetic fields
enhancing the polarized emission \citep[e.g.][]{ensslin98,kang12,kang13}.  

The Coma cluster is one of the nearest galaxy clusters and the 
first where both a central radio halo \citep{large59} and a peripheral radio relic
\citep{jaffe79, ballarati81} have been detected. 
During the past four decades, these two sources have been extensively
analyzed at several radio frequencies, allowing a good determination of their radio spectra up to $\simeq$ 5 GHz \citep[e.g.][]{andernach84,thierbach03}.
Deep low frequency observations have shown that the halo and the relic of the Coma galaxy clusters are connected by a bridge of diffuse radio emission \citep[e.g.][]{kim89,venturi90,brown11,bonafede21}. At 144\,MHz, the radio halo extends for more than 2\,Mpc \citep{bonafede22}. The properties of the Coma radio halo can be
explained by models invoking diffuse reacelleration of seed relativistic electrons \citep[e.g.][]{brunetti01,brunetti07,marchegiani19}.
The presence of even larger-scale diffuse emission around Coma has been suggested by \citet{kronberg07}, who detected patches of distributed  radio “glow” beyond the scale of halo, relic, and bridge. 

The study of the spectra of radio halos and relics is particularly
interesting at high frequencies, where important constraints on the physical
mechanisms responsible for their formation and evolution can be found \citep[e.g.][]{loi20,  rajpurohit20}. 
However, accurate measurement of their 
integrated spectra over a broad range of frequencies is a difficult task.
Firstly, they are usually embedded in a number of discrete sources, whose 
flux densities need to be carefully subtracted from the total diffuse emission.
Secondly, due to their steep spectra, these sources have a low surface brightness at GHz frequencies.
Finally, high-frequency observations of extended emission are limited by technical difficulties.
Interferometers suffer the so-called missing zero spacing problem, where structures larger than the angular size
corresponding to the shortest spacings, are absent, thus making them “blind”
to extended structures, while maintaining the high resolution.
On the other hand, single dish observations are
optimal to detect all the emission up to a scale equal to the size of the scanned region, but will lack emission on
angular scales smaller than the primary beam size.
One possible solution to these problems is to combine interferometric data with
single dish ones to obtain final images with both the high resolution of
the interferometer and the short-spacing information 
provided by the single dish.

To study in detail the high frequency properties of the diffuse radio emission in the Coma cluster,
in this paper we present new Sardinia Radio Telescope \cite[SRT;][]{bolli15,prandoni17}
observations in C-Band (6.6 GHz) and L-Band (1.4 GHz). We combined L-Band SRT observations  
with Very Large Array (VLA) interferometric observations to obtain a final image with both the high resolution of the interferometer and the short-spacing information provided by SRT.
We observed the Coma cluster in L and C bands as part of the SRT Multi-frequency Observations of Galaxy Clusters (SMOG) project.
The SMOG project (PI M. Murgia; SRT Project S0001) is an SRT early science program focusing on wideband and wide-field single dish spectral-polarimetric studies of a sample of galaxy clusters.
The aim of this project is to use SRT observations 
of a sample of nearby galaxy clusters known from the literature to harbor diffuse radio halos, 
relics, or extended radio galaxies to investigate the intra-cluster media on  large scales and to 
understand the interplay between relativistic particles and magnetic fields
and the life-cycles of cluster radio galaxies \citep{murgia16,govoni17,loi17,vacca18,brienza18}.
We also added to the SMOG data of the Coma cluster the C-Band observations collected during the commissioning of the SARDARA 
(SArdinia Roach2-based Digital Architecture for Radio Astronomy)
backend \citep{melis18}.

We describe the radio observations and the data reduction in Sect. 2. We present the resulting total intensity and the polarized diffuse emission of the Coma cluster in Sect. 3. The analysis of the halo, and comparison with literature data, is presented in Sect. 4. The analysis of the relic, and comparison with literature data, is presented in Sect. 5.
In Sect. 6, we discuss our findings and give the conclusions in Sect. 7. In Appendix A, we 
provide details of the imaging stacking algorithm used in the SRT data reduction. In Appendix B, we report a consistency check of the SRT flux density scale. In Appendix C, we look for possible counterparts the halo polarized spot at different frequencies. In Appendix D, we report the parameters of the SRT beam at 6.6\,GHz.

Throughout this paper we assume a $\mathrm{\Lambda CDM}$ cosmology with
$\mathrm{H_0= 71 km\, s^{-1}\, Mpc^{-1}}$,
$\mathrm{\Omega_m=0.27}$, and $\mathrm{\Omega_{\Lambda}=0.73}$.
At the Coma redshift \cite[$z=0.0231$;][]{struble99}, 1 arcsec\, corresponds to 0.46\,kpc.

\section{Radio observations and data reduction}

The C-band observations were taken between 2015 May and 2016 July, with a total of 
about 90 hours on source. The L-Band observations were taken in 2016 July, with a total of 14 hours on source. A summary of the SRT observations is shown in
Table\,\ref{tab:SRT_observations}. The data reduction and imaging were done 
with the Single-dish Spectral-polarimetry Software \cite[SCUBE;][]{murgia16}.

\begin{table*}
	\centering
	\caption{Details of the SRT observations centered on $RA_{\rm J2000}=12^{\rm h}57^{\rm m}17^{\rm s};DEC_{\rm J2000}=+27^{\circ}47\arcmin49\arcsec$.
	 Col. 1: Frequency range; Col. 2: Spectral resolution; Col. 3: Observing time (calibrators included); Col. 4: Date of observation;  
	 Col. 5: Number of images on the source; Col. 6: Calibrators; Col. 7: SRT project name.
	}
	\label{tab:SRT_observations}
	\begin{tabular}{|ccclcll|} 
	  \hline
Frequency Range      & Spectral Resolution&  Time On Source  & Observing Date     & OTF Mapping     & Calibrators & SRT Project  \\	  
(MHz)            & (MHz/channel)  &    (hours)      &                    &                 &             &            \\ 
\hline
C-Band           &            &                 &               &                 &             &             \\ 
6000$-$7200      & 1.46        &      3         & 2015 May 2   & 1RA$\times$1DEC &  3C\,286, 3C\,84     & SARDARA comm.  \\
                    &         &      3         & 2015 Jul 3  & 1RA$\times$1DEC &  3C\,286, PKS1830-21 & SARDARA comm.  \\
                    &         &      3         & 2015 Jul 16  & 1RA$\times$1DEC &  3C\,286, PKS1830-21 & SARDARA comm.  \\
                    &         &      3         & 2015 Jul 31  & 1RA$\times$1DEC &  3C\,286, PKS1830-21 & SARDARA comm.   \\
                    &         &      3         & 2015 Aug 7  & 1RA$\times$1DEC &  3C\,286, PKS1830-21 & SARDARA comm.   \\
                    &         &      9         & 2015 Dec 5  & 3RA$\times$3DEC &  3C\,286, 3C\,84     & SARDARA comm.  \\
                    &         &      3          & 2016 Feb 2/3 & 1RA$\times$1DEC &  3C\,286, 3C\,84  &   S0001 \\
                    &         &      9          & 2016 Feb 6/7& 3RA$\times$3DEC &  3C\,286, 3C\,84  &   S0001  \\
                    &         &      7.5          & 2016 Mar 14/15 & 2RA$\times$3DEC &  3C\,286, 3C\,84  &   S0001  \\
                    &         &      7.5          & 2016 Mar 15/16& 3RA$\times$2DEC &  3C\,286, 3C\,84  &   S0001  \\
                    &         &      7.5          & 2016 Mar 17/18 & 2RA$\times$2DEC &  3C\,286, 3C\,84  &   S0001  \\
                    &         &      7.5          & 2016 Mar 18/19& 3RA$\times$2DEC &  3C\,286, 3C\,84  &   S0001  \\
                    &         &      7.5          & 2016 Mar 19/20 & 3RA$\times$2DEC &  3C\,286, 3C\,84  &   S0001  \\
                    &         &      7.5          & 2016 Mar 20/21& 2RA$\times$3DEC &  3C\,286, 3C\,84  &   S0001  \\
                    &         &      7.5          & 2016 Jul 27& 2RA$\times$2DEC &  3C\,286, 3C\,84  &   S0001  \\

\hline
L-Band           &            &                 &                 &                 &                   &         \\ 
1300$-$1800      & 0.0916         &      7          &  2016 Jul 08/09 & 3RA$\times$3DEC &  3C\,147, 3C\,286,           &   S0001  \\
                 &               &                 &                 &                   &   3C\,84, 3C\,138                            &           \\  
                 &            &      7          &  2016 Jul 18   & 3RA$\times$3DEC &  3C\,147, 3C\,286,          &   S0001  \\
                 &             &                &                 &                 & 3C\,84, 3C\,138            &          \\
\hline
	\end{tabular}
\end{table*}

\subsection{C-Band observations}
We observed with the SRT an area of $2\,{\rm deg}\times 2\,{\rm deg}$ in the field of the Coma cluster centered at $RA_{\rm J2000}=12^{\rm h}57^{\rm m}17^{\rm s}$ and $DEC_{\rm J2000}=+27^{\circ}47\arcmin49\arcsec$, using the C-Band receiver \citep{peverini09}. Full-Stokes parameters were acquired with the SARDARA backend in the frequency
range 6000$-$7200\,MHz, at a spectral resolution of 1.46\,MHz per channel. 
We performed several on-the-fly \cite[OTF;][]{mangum07} mapping in the equatorial frame, 
alternating the RA and DEC directions. 
The telescope scanning speed was set to 6\,arcmin/sec and the scans were 
separated by 0.7\,arcmin to properly sample the SRT beam,
whose FWHM is 2.9\,arcmin in this frequency range. 
We recorded the data stream by acquiring 33 full bandwidth spectra per second. Individual samples (each consisting of an independent spectrum) are spaced across the sky by 11\,arcsec along the scan direction. Using this set-up, the beam FWHM is sampled with four cells in the direction orthogonal to the scan direction, and is greatly over sampled along the scan direction, but this is averaged out in the final pixelization.
Band-pass, flux density, and polarization calibrators were done
with at least four cross scans on source calibrators performed
at the beginning and at the end of each observing run.

\begin{figure*}
    \centering
    \includegraphics[width=0.95\textwidth,trim=15 260 140 15]{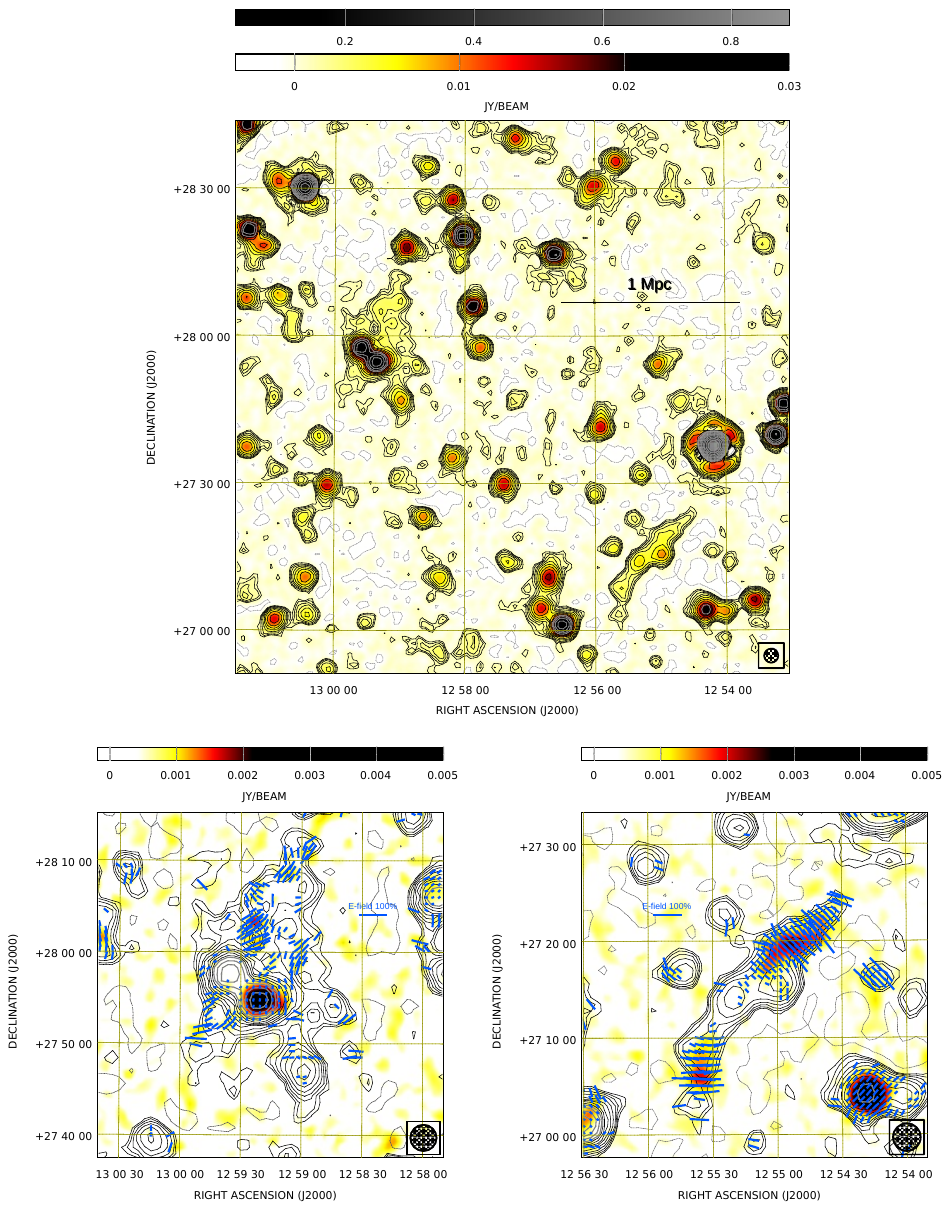}
    \caption{\emph{Top panel:} SRT total intensity image at 6.6\,GHz of the Coma cluster, resulting from the average of all spectral channels in the 1.2\,GHz bandwidth. Contours start at 3$\sigma_{I}$ where $\sigma_{I}$=0.33\,mJy\,beam$^{-1}$ and increment with a $\sqrt{2}$ factor. Dotted light-gray contours are negative contours drawn at -3$\sigma_{I}$. The beam FWHM is 2.9\,arcmin, as shown in the bottom right corner of the images. \emph{Bottom panels:} SRT linearly polarized intensity image for the radio halo (left) and relic (right) obtained from the spectral average. The polarization noise level, after the correction for the positive bias, is $\sigma_{P}$=0.25\,mJy\,beam$^{-1}$. Contours refer to total intensity, while vectors represent the orientation of the polarization angle and are proportional in length to the fractional polarization $\mathrm{FPOL}  =P/I$. Polarization vectors refer to the radio wave $\Vec{E}$-field and are traced only where $\mathrm{FPOL}   > 3\sigma_{\mathrm{FPOL}  }$. The reference bar corresponds to a 100\% polarized signal.}
    
    \label{coma_6.6}
\end{figure*}

\begin{figure*} 

    \includegraphics[width=0.95\linewidth,trim=25 510 0 0]{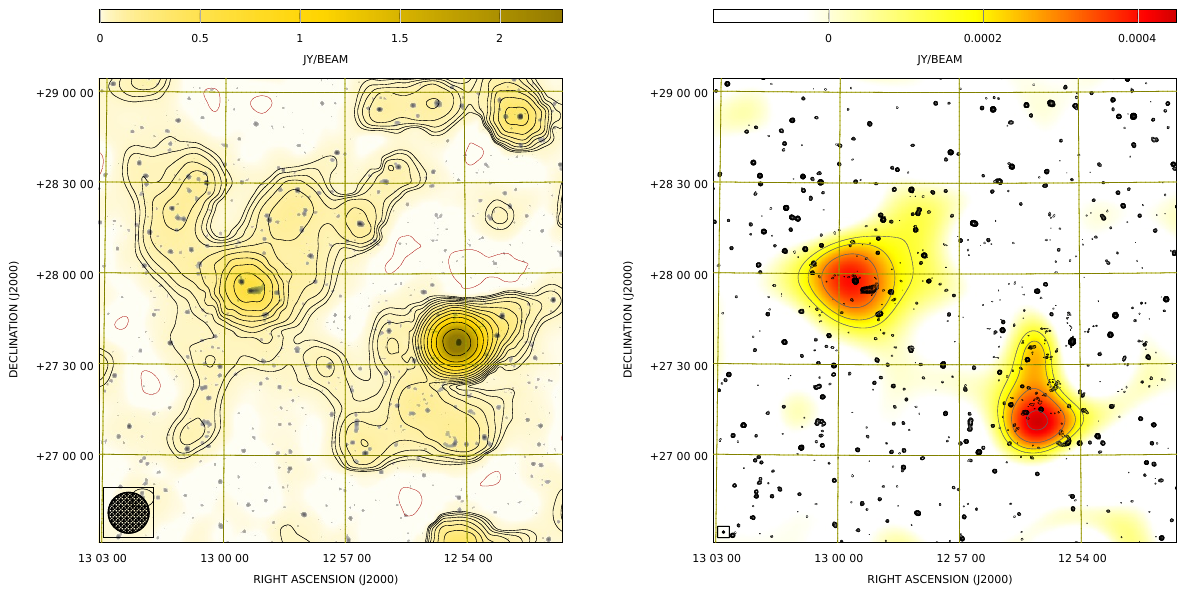}
    \includegraphics[width=0.95\textwidth,trim=50 570 60 25]{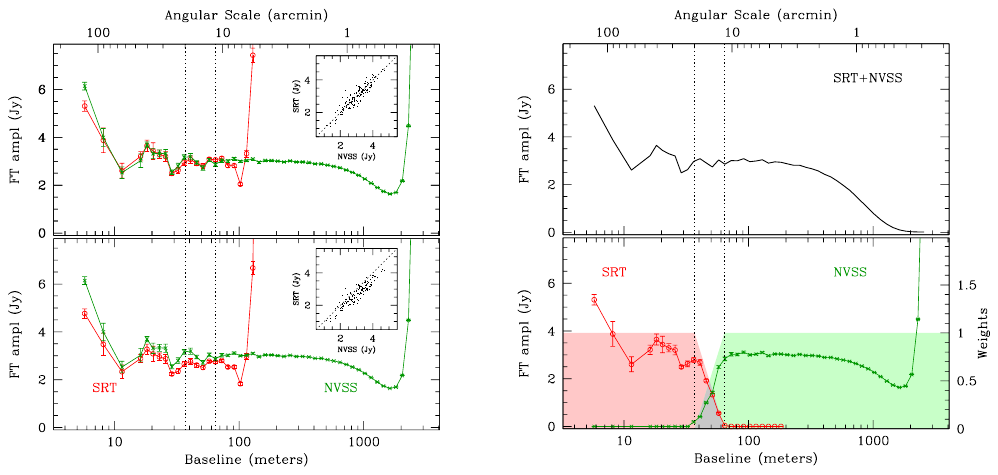}

    \caption{\emph{Top-left:} SRT image of the Coma cluster at 1.4\,GHz represented in contours and yellow tones over the NVSS image (grayscale). Contours start at 30 mJy/beam (confusion limit) and increase by $\sqrt{2}$. The beam FWHM is 13.5\,arcmin.  \emph{Top-right:} Black contours refer to the combination of the SRT and NVSS images after their merging in the Fourier space. Levels start at 1.5 mJy/beam (45\,arcsec beam) and increase by $\sqrt{2}$. The color image and the gray contours represent the contribution to the extended emission due to the SRT data only, with levels starting at $3\sigma$, or 0.15 mJy/beam (45\,arcsec beam), and increasing by $\sqrt{2}$ (see text). \emph{Bottom left:} Azimuthally averaged radial profiles of the Fourier amplitudes for the SRT (red dots) and the NVSS (green crosses). The vertical dotted lines delimit the ring in the Fourier space, where the SRT and the NVSS spatial scales overlap (baselines from 35 to 64\,m). The lower panel shows the initial amplitudes of the Fourier transforms deconvolved from the respective observing beam. The SRT amplitude needs to be increased by a factor of 11\% to match the NVSS flux density. The upper panel shows the Fourier amplitude profiles after the application of the scaling 
    factor to the SRT data. The insets show the consistency of the NVSS and SRT flux density inside the ring. \emph{Bottom right:} The lower panel shows the linear weighting of the interferometer and single-dish amplitudes, as represented by the shaded regions and the right-axis scale.
    The upper panel shows the final merged Fourier amplitude re-convolved to the 45\,arcsec angular resolution of the NVSS.}
    \label{fig:srt_over_nvss}
\end{figure*}

\begin{figure*}
  \includegraphics[width=0.95\linewidth,trim=40 650 40 0]{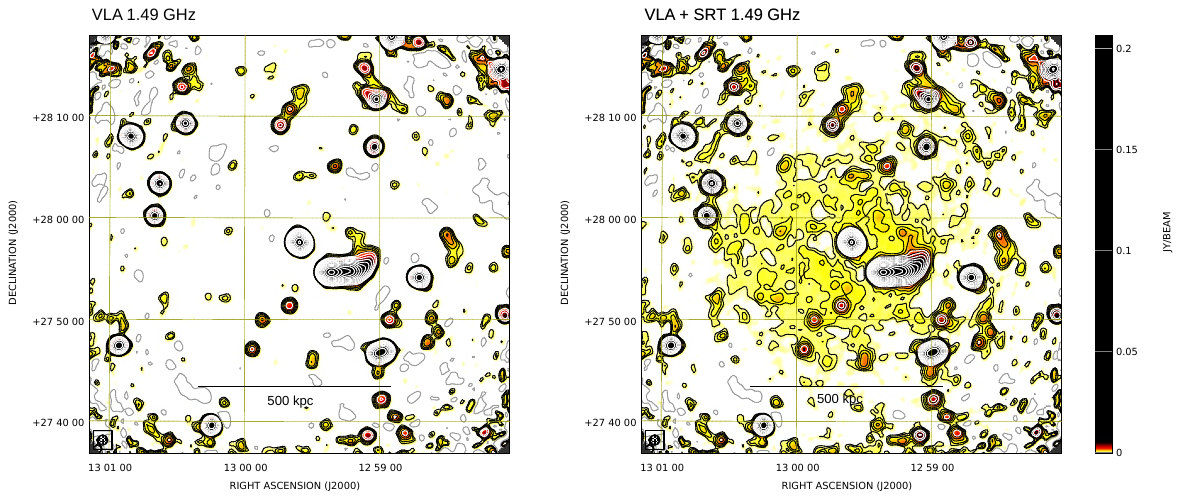}

    \includegraphics[width=0.5\textwidth,trim=0 175 0 0]{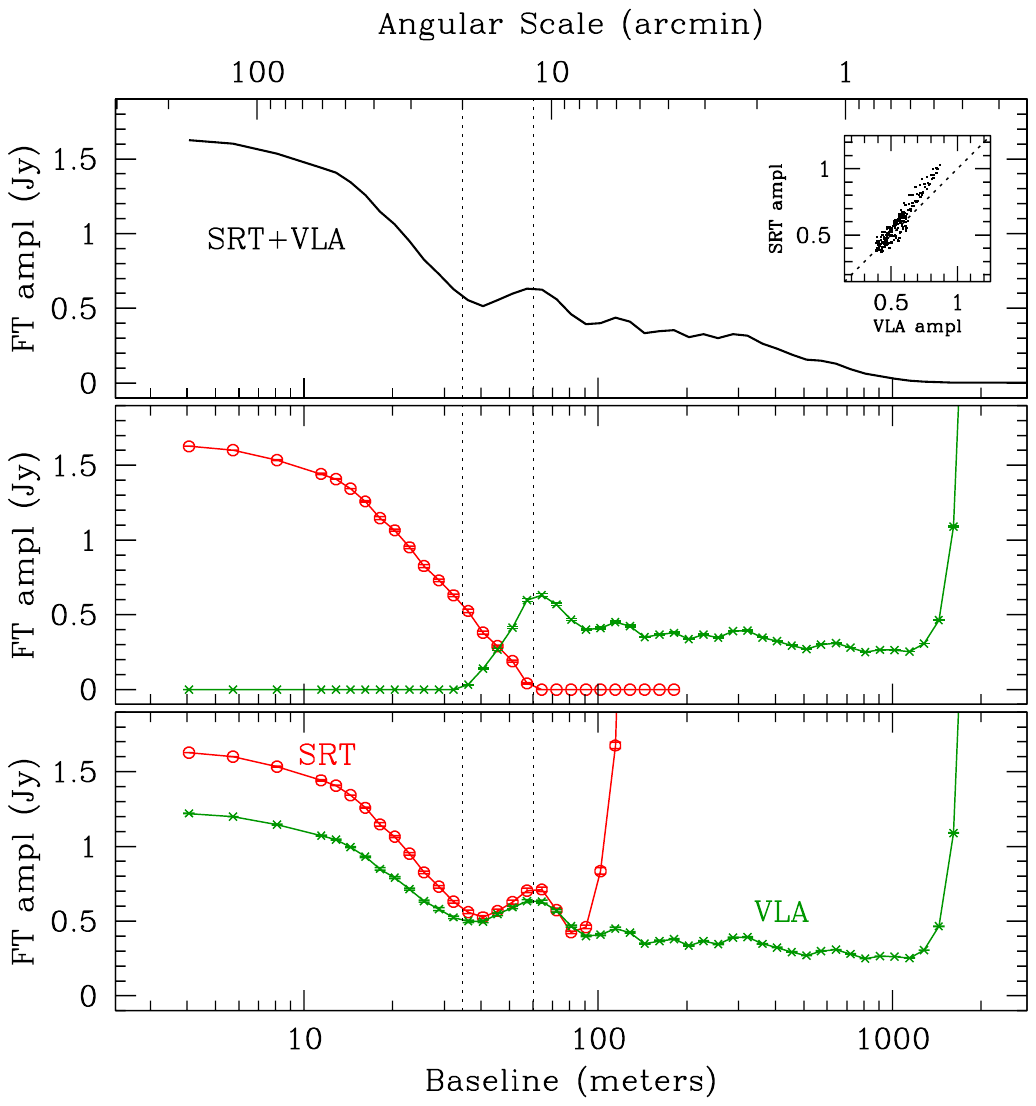}

    \caption{Radio images at 1.49\,GHz of the center of the Coma cluster. \emph{Top-left:} VLA D-configuration image. \emph{Top-right:} Combination of the SRT and VLA images. In both images, contours start a 3$\sigma_{I}=0.45$ mJy/beam and increase by $\sqrt{2}$. The beam FWHM is 60\,arcsec. One negative level at -3$\sigma_{I}$ is represented in light-gray. \emph{Bottom:} plot of the Fourier amplitude as a function of baseline in meters. The lower panel shows the SRT and VLA amplitude deconvolved by the respective point spread functions. The middle panel shows the weighted profiles of the single-dish and interferometer Fourier amplitudes, with a scaling factor of 11\% applied to the SRT amplitude. The upper panel shows the combined amplitude re-convolved to the VLA beam of 60\,arcsec. The inset shows the consistency of the VLA and SRT flux densities inside the overlapping annulus before the final weighted merging.}
    \label{srt_over_vlaD_halo}
\end{figure*}

We used a standardized pipeline for the calibration and imaging.
We excised all the RFI at well known specific frequencies. 
Band-pass and flux density calibration were performed by
observing 3C\,286, assuming the flux density scale of \citet{perley17}. 
After a first bandpass and flux density calibration cycle, we 
flagged spectral channels affected by persistent RFI.
We then repeated a second round of bandpass and the flux density calibration.
We applied the gain-elevation curve correction to account for the
gain variation with elevation due to the telescope structure
gravitational stress change. We performed the polarization calibration 
by correcting the instrumental on-axis
polarization and the absolute polarization angle.
The on-axis instrumental polarization was determined through observations
of the bright unpolarized sources 3C\,84 or PKS\,1830-21. 
The leakage of Stokes $I$ into $Q$ and $U$ is in general less than 2\% across
the band, with an RMS scatter of $0.7 - 0.8$\%.
We fixed the absolute position of the polarization angle using as reference
the primary 
polarization calibrator 3C\,286.
The difference between the observed and predicted position angle according to \citet{perley13}  
was determined, and corrected channel-by-channel.
The calibrated position angle was within the expected value 
of 33$^{\circ}$ for 3C\,286, with a RMS scatter of $\pm$ 1$^{\circ}$.

Our primary goal is to image the emission of the diffuse emission and the discrete
sources in the field-of-view of the Coma cluster. Since we are not interested in retaining any large scale foreground emission, we removed the baseline scan-by-scan by fitting a 2nd-order polynomial to the “cold-sky” regions devoid of both discrete sources and of 
the galaxy cluster extended emission (relic and halo). These cold-sky regions were 
identified using a mask created with the NRAO VLA Sky Survey 
\cite[NVSS;][]{condon98} image at 1.4\,GHz, convolved with a beam FWHM 
of 2.9\,arcmin and including those sources with a brightness $>$0.5 mJy/beam. 
In this way, we removed the base level related to the receiver noise, the atmospheric emission, and the large scale foreground sky emission. 
We then imaged the spectral cubes of the $L$, $R$, $U$ and $Q$ polarizations of all the observing slots. To increase the sensitivity, for each polarization, we stacked all spectral cubes separately for the RA and DEC directions. We used a weighted stacking to take into account of the different noises of the scans and to taper the region affected by residual RFI, see Appendix \ref{appendixA} for details.  

Finally, we combined all the RA and DEC scans by mixing their stationary wavelet transform (SWT) coefficients \citep[see][]{murgia16}. The de-striping resulting from the SWT stacking is effective in isolating and removing the characteristic noise oriented along the scanning direction. 
\begin{figure*}
  \includegraphics[width=0.95\linewidth,trim=40 650 40 0]{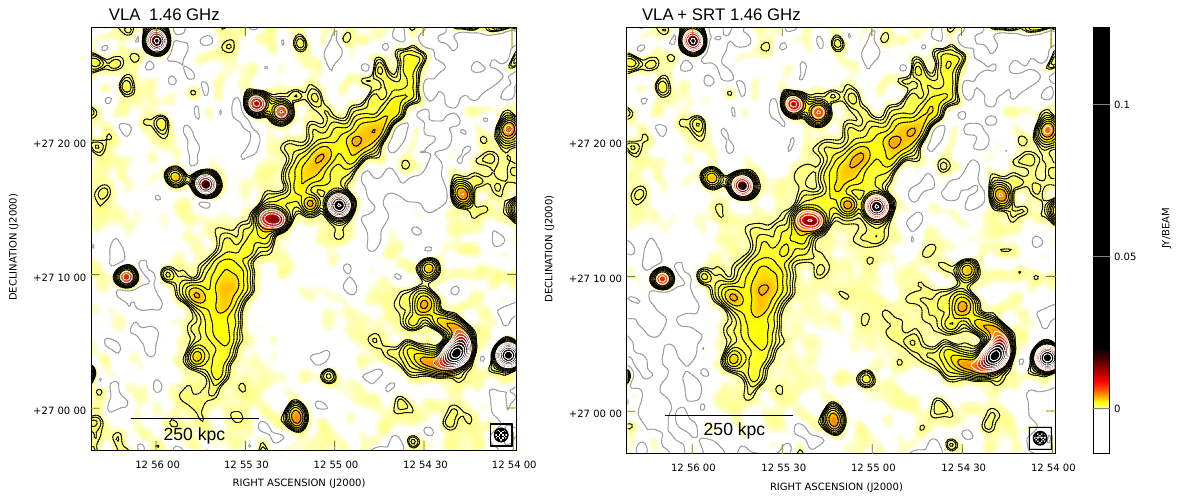}
    \includegraphics[width=0.5\textwidth,trim=0 175 0 0]{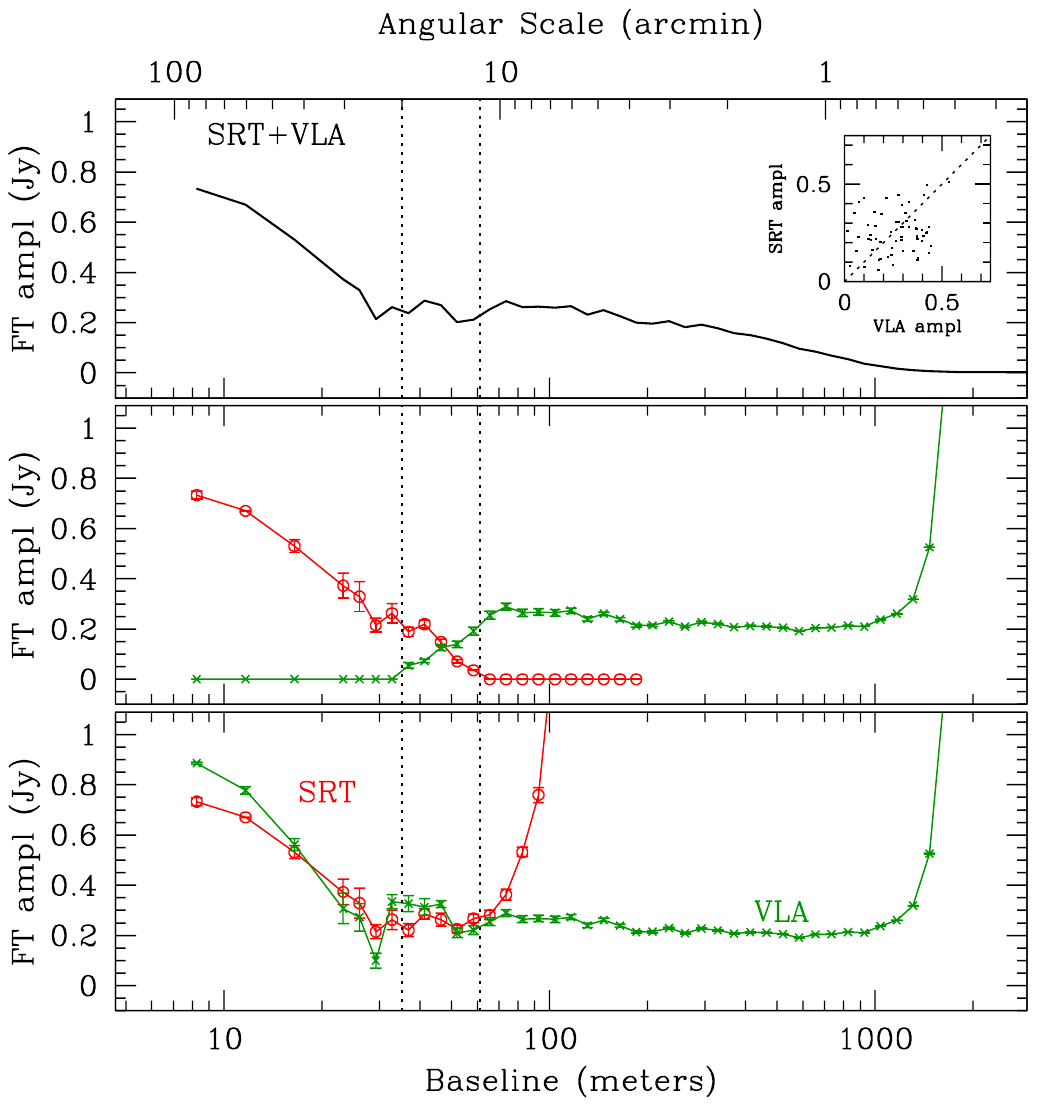}

    \caption{Radio images at 1.46\,GHz of the peripheral relic  B1253+275. \emph{Top-left:} VLA D-configuration image. \emph{Top-right:} Combination of the SRT and VLA images. In both images, contours start a 3$\sigma_{I}=0.45$ mJy/beam and increase by $\sqrt{2}$. The beam FWHM is 60\,arcsec. One negative level at -3$\sigma_{I}$ is represented in light-gray. \emph{Bottom:} plot of the Fourier amplitude as a function of baseline in meters. In the lower panel shows the SRT and VLA amplitude deconvolved by the respective point spread functions. Middle panels show the weighted profiles of the single-dish and interferometer Fourier amplitudes, with a scaling factor of 11\% applied to the SRT amplitude. The upper panels show the combined amplitude re-convolved to the VLA beam of 60\,arcsec. The inset shows the consistency of the VLA and SRT flux densities inside the overlapping annulus before the final weighted merging.}
    \label{srt_over_vlaD_relic}
\end{figure*}

The images of the total intensity $I$, polarized intensity $P$, and the polarization angle $\Psi$ have been derived from the circular polarizations $R$ and $L$, and from Stokes parameters $U$ and $Q$, according to:
\begin{eqnarray}
   I & = & R+L,\\ 
   P & = & \sqrt{U^2+Q^2},\\ 
    \Psi & = & 0.5 \cdot \arctan(U/Q).
\end{eqnarray}
We also corrected the polarization images for the Rician bias \citep{killeen86}. 

\subsection{L-Band observations}

We observed with the SRT an area of $3\,{\rm deg}\times 3\,{\rm deg}$ centered 
at  $RA_{\rm J2000}=12^h57^m17^s$ and $DEC_{\rm J2000}=+27^{\circ}47\arcmin49\arcsec$ in the field of the Coma cluster
using the L-and P-band dual-frequency coaxial receiver for 
primary-focus operations \citep{valente2022}. We used the L-band component only covering the
frequency range 1.3-1.8\,GHz. The data stream was recorded with the SARDARA backend.
The configuration used had 1500 MHz of bandwidth and 
16384 channels $\sim$ 91.6\,kHz each, full-Stokes parameters, and
sampling at 10 spectra per second. We opted for the OTF mapping strategy scanning the sky alternatively
along the RA and DEC direction with a telescope scanning
speed of 6\,arcmin/sec and an angular separation of 3\,arcmin between scan. This choice guarantees a proper sampling of the SRT beam
whose full width at half maximum (FWHM) is about 14\,arcmin at 1550 MHz, the central frequency of these observations.
Since the sampling time is 100 ms, the individual samples
are separated in the sky by 36\,arcsec along the scanning direction, but the uneven beam sampling will be compensated for by averaging in the final pixelization.

The SRT L-band receiver is native linear, and we acquire the vertical and horizontal polarizations, $X$ and $Y$, and the Stokes parameters $U$ and $V$. We used 3C\,147 as band-pass and flux density calibrator and the flux density scale of \citet{perley17} was assumed. 

The RFI contamination in the L-band is substantial, and it is necessary to refine 
the data reduction by iterating RFI flagging and calibration. 
The pipeline starts by flagging all spectral regions known to be affected by the presence of persistent RFI.
This is done using a fixed flag table. An automatic flag algorithm is then used to search 
both along time and frequency for any other RFI not included in the
initial flag table. An initial set of band pass and flux density scale solutions are then obtained. Using this first calibration, we flagged all calibrators' data outside a specific flux density range (from -1 to 15 Jy) and we repeated the entire procedure to obtain refined bandpass and flux density solutions. At the end of the process, about 25\% of the data were flagged.

We ran the automatic flag algorithm in the calibrated data of the Coma cluster. A “cold-sky” model for the Coma field is obtained using the NVSS, smoothed at SRT resolution. We masked out all point sources above an intensity level of 2$\sigma$, the halo, the bridge, and the relic regions, and we used the remaining portion of the field of view to model the subscan baseline with a first-order polynomial (linear fit). We subtracted 
the baseline from the data to remove receiver noise, atmospheric contributions, and all unwanted sky emission on scales larger than our FoV. We then flagged all data outside a specific flux density range (from -0.2 to 2 Jy) and we repeated the procedure to obtain a refined baseline subtraction. 
Finally, the baseline-subtracted data were then projected onto a regular three-dimensional grid with a spatial resolution of 3\,arcmin/pixel. These steps were performed in each frequency channel for both the $X$ and $Y$ polarizations. We calibrated and analyzed only the total intensity of the Coma cluster at L-band.

The frequency cubes for $X$ and $Y$ polarizations were stacked separately for RA and DEC scans, following the same weighted average technique described in the Appendix\,\ref{appendixA}.  As for the C-band observations, the final RA and DEC cubes were merged and de-striped by mixing the SWT coefficients. 

As a last step, the total intensity cube, $I$, was formed by summing the two polarizations: $I=X+Y$.

\begin{figure*}
    \centering
    \includegraphics[width=0.95\textwidth,trim=50 310 50 15]{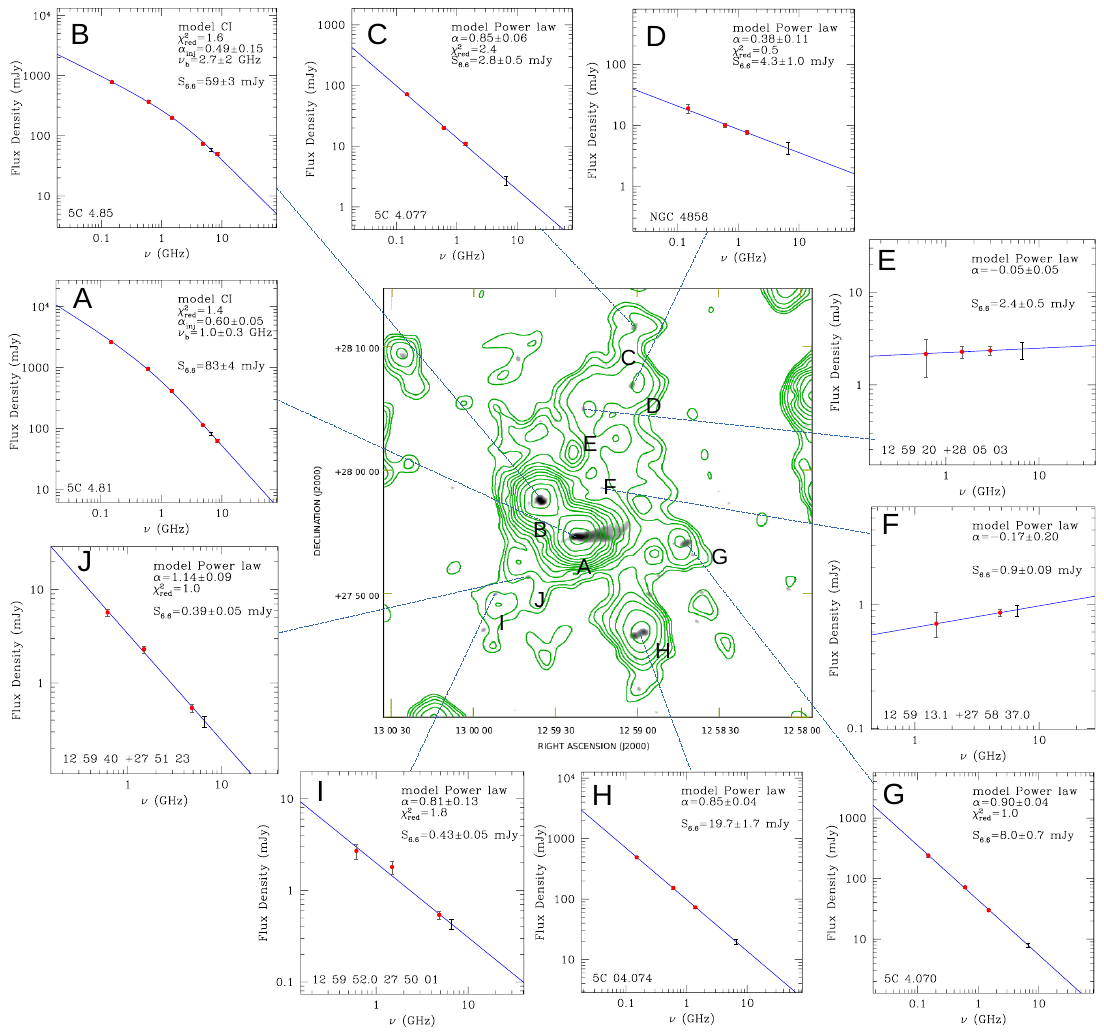}

    \caption{VLA C-configuration image at 1.49 GHz and 15\,arcsec resolution with the SRT radio contours at 6.6\, GHz overlaid. The panels show the spectra of the ten brightest discrete radio sources embedded in the radio halo. The bar shown in each spectrum marks the expected flux density at 6.6\,GHz. For the two strongest sources at the center of the cluster, 5C\,4.81 and 5C\,4.84, the solid line represents the best fit of the CI model. For the remaining sources, the solid line represents the best fit of a power law. 
    }
    \label{fig:halo_ps}
\end{figure*}

\begin{figure}
    \centering
    \includegraphics[width=1.0\columnwidth,trim=25 200 25 175] {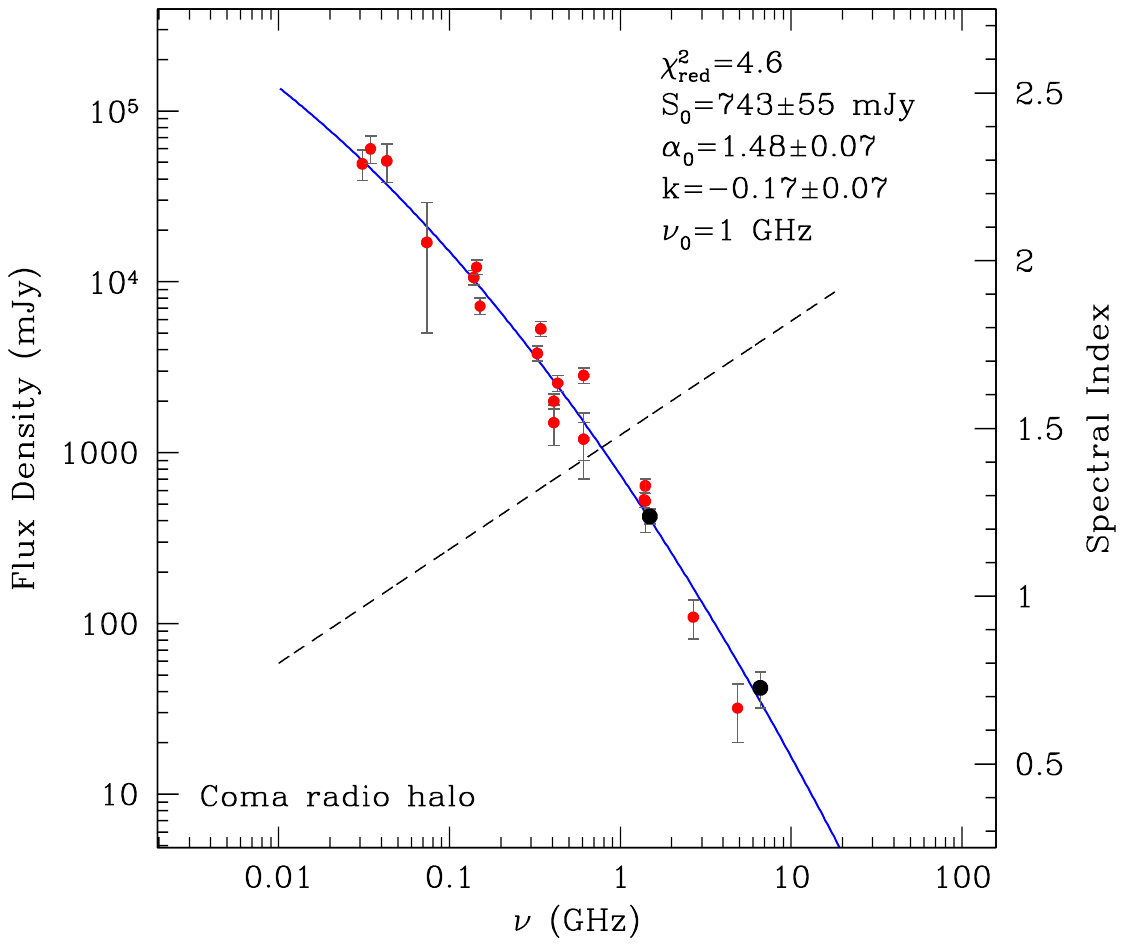}
    \caption{Global radio spectrum of the Coma halo. Black dots are measurements from this work, while red dots are the measurements from the literature. The modified power law best-fit is shown as a solid blue line. The dashed line represents the model spectral index as function of frequency (left axis scale).}
    \label{halo_global_spectrum}
\end{figure}

\section{Results}

\subsection{C-band total intensity and polarization images}

In the top panel of Fig. \ref{coma_6.6} we show the 6.6\,GHz SRT total intensity image resulting from the average of all spectral channels in the 1.2\,GHz bandwidth. The beam FWHM is 2.9\,arcmin. We estimated the image noise level using 100 circular regions of exactly one beam in size scattered all over the background in the field of view (FoV). We calculated the dispersion of the average intensity between all the boxes, and found an RMS noise of $\sigma_{I}=$0.33\,mJy/beam. This value is very close to the expected confusion noise limit of $\sigma_c=0.36$\,mJy/beam expected at this frequency and angular resolution, for a spectral index $\alpha=0.8$ \citep{condon02}. We draw solid contours from 3$\sigma_{I}$ increasing by a factor of $\sqrt{2}$, while light-gray contours are the negative -3$\sigma_{I}$ contours.  We clearly detect the emission associated to the brightest radio sources in the field, part of the diffuse radio halo (Coma C) at the cluster center and the peripheral radio relic (B1253+275) located on the southwest of the cluster center. The detection of the central radio halo and the peripheral relic at 6.6\,GHz presented in this work are the images at the highest frequency ever obtained for these diffuse sources in the Coma cluster.

In the SRT 6.6\,GHz image, the radio halo diffuse emission is elongated in the north-south direction and the largest linear size, measured from the $3\sigma_I$ isophote, is approximately 30\,arcmin, corresponding to 830\,kpc. The overall shape and elongation of the radio halo are broadly consistent with the images at 2.675 and 4.85\,GHz obtained at a similar resolution with the Effelsberg 100-m telescope by \citet{thierbach03}. However, the elongation of the halo toward north-west is due in part to a series of point sources incorporated in the diffuse emission, and is analyzed in detail in Sect.\,\ref{radio_analysis_6.6}.
The radio relic has an elongated structure with a fairly uniform brightness, which extends over a linear length of 770\,kpc (or 28\,arcmin in angular size) with a deconvolved width of $\sim 80$\,kpc.

We checked the consistency of the calibration of the flux density scale using Coma-A (alias 3C\,277.3; RA: 12$^h$54$^m$12$^s$, DEC: +27$^d$37$^m$34$^s$), the strongest radio source in the FoV. Coma-A is an ideal radio source to cross-check the calibration since its emission is dominated by two radio lobes and, given its angular size of $LAS\simeq 1\arcmin$, is unresolved by the SRT at 6.6\,GHz. By considering all the fifteen observing sessions, the systematic uncertainty for the SRT flux density measurements at 6.6\,GHz is estimated to be about 1\%. See Appendix\,\ref{appendixB} for details.

In the bottom-left panel of Fig. \ref{coma_6.6} we show a zoom of the cluster center where we have overlaid the radio halo total intensity contours on the 6.6\,GHz SRT polarized intensity image obtained by the average of all spectral channels in the 1.2\,GHz bandwidth. The polarization noise level, after correction for the positive bias, is $\sigma_{P}$=0.25\,mJy\,beam$^{-1}$. We show the polarization vectors of the $\vec{E}$-field where the fractional polarization, defined as $\mathrm{FPOL}=P/I$, is detected with S/N>3. 
We did not correct the polarization vectors for the Rotation Measure effect. The Coma cluster Galactic coordinates are $l=58^{\circ}$ and
$b=88^{\circ}$. The cluster is close to the galactic North Pole and the galactic RM is so low, $\sim 0.3$\,rad/m$^2$ \citep{oppermann11}, that its effect on the observed polarization angle is completely negligible at 6.6\,GHz. The bright polarized intensity peak, close to the center of the halo, is due to the head-tail source 5C\,4.81 which is nearly point-like to  the SRT beam. The fractional polarization of 5C\,4.81 is $\mathrm{FPOL} \simeq 5$\% at 6.6\,GHz. This fractional polarization level is significantly less than that measured by \citet{Bonafede10} at 1.5\,arcsec resolution, most likely because of the significant beam depolarization affecting the SRT image. The close-by source to the north-east, 5C\,4.85, is completely unpolarized in the SRT image for the same reason. With the VLA at 1.5\,arcsec resolution, \citet{Bonafede10} measured fractional polarization levels of 18\% and 10\% for 5C\,4.81 and 5C\,4.85, respectively.

A localized spot of polarization is observed in the radio halo at RA: 12$^h$59$^m$24$^s$, DEC: +28$^d$03$^m$00$^s$ with an intensity $I=2$ mJy/beam and $P=0.9$ mJy/beam ($\mathrm{FPOL}  \simeq 45\%$). There is a hint that the polarized emission extends along the north-east side of the halo at a level of $\mathrm{FPOL}  \simeq 35\%$. We checked for the presence of a discrete source (such as a fossil steep spectrum source) at the position of the spot that could explain the observed level of polarization. We inspected the LOFAR \citep[LoTSS;][]{shimwell22} and GMRT \citep[TGSS;][]{intema17} images 
 at 144 and 150\,MHz, but we do not find the presence of any discrete radio source, embedded in the halo diffuse emission in this position. We also verified in the VLA Sky Survey \citep[VLASS;][]{lacy20}  at 2-4\,GHz for a possible source visible only at high frequency due to of an inverted radio spectrum, but we find none (see Appendix\,\ref{appendixC}).
 
 Although many radio halos have been imaged well in total intensity, their polarized signals are still very difficult to detect, and so far, only a few radio halos have been imaged in polarization. Using Magneto-Hydro-Dynamical cosmological simulations, \citet{govoni13} demonstrated that most radio halos could be intrinsically polarized, but the resolution and sensitivity of current instruments hinder the detection of this signal. In particular, for the Coma cluster, where
the power spectrum of intracluster magnetic field fluctuations extends at most to a maximum scale of a few tens of kpc \citep{Bonafede10}, we do not expect large-scale polarization associated with the radio halo.
 However, simulations have shown that, even if most of the polarized emission drops below the noise level, localized spots
could be detectable in Coma-like halos (see e.g. Fig.\,3 in \citealt{govoni13}).

 In the bottom-right panel of Fig. \ref{coma_6.6} we show a zoom of the radio relic region with the total intensity contours overlaid on the 6.6\,GHz SRT polarized intensity image.  The radio relic is highly polarized, from $\mathrm{FPOL}  \simeq 40\%$ in the north-west tip up to $\mathrm{FPOL}  \simeq 50\%$ in the south-east tip. The polarization vectors are perpendicular to the relic ridge line, as typically observed for these radio sources. At 2.675\,GHz and 4.3\,arcmin resolution, \citet{thierbach03} observed the same  polarization angle distribution but slightly lower fractional polarization levels of $\mathrm{FPOL}  \simeq 30\%$ in the north and  $\mathrm{FPOL}  \simeq 40\%$ in the south of the relic, respectively. The fractional polarization levels they measured are in good agreement with those measured with the SRT when considering the beam depolarization effect that could be caused by the coarser angular resolution of their observation.

\subsection{L-band total intensity images}

In the top-left panel of Fig.\,\ref{fig:srt_over_nvss} we present the 1.4\,GHz SRT total intensity image, represented by contour levels and yellow tones, overlaid on the NVSS image, represented by gray tones. The SRT image is obtained by averaging all spectral channels in a bandwidth of 200\,MHz centered at 1.4\,GHz. The FWHM beam size is 13.5\,arcmin and the image is confusion limited with $\sigma_{c}\simeq 30$ mJy/beam.  

We combined our single-dish SRT data together with interferometric data to improve the angular resolution while maintaining the sensitivity to large-scale structures present in the field of view. We computed the combination in the Fourier transform plane using task \textsc{isd\_combo} in SCUBE.

As a first step, we perform a fine calibration of the SRT flux density scale in the Fourier space to match that of the NVSS in the spatial scale interval common to the two images, as illustrated in the bottom-left plot of Fig.\ref{fig:srt_over_nvss}. Since processing of the single-dish scans involves fitting a baseline level, this data set may be the one with the most uncertain calibration and a multiplicative factor is used for a finer correction of image values. The same single-dish scaling factor is implemented in the analogous tasks AIPS \textsc{imerg} and MIRIAD \textsc{immerge}.  The NVSS is effective to determine the scaling factor of the SRT image, given the large number of point sources contained in the mosaic. The plots represent the azimuthally averaged radial profiles of the Fourier amplitudes for the SRT (red dots) and the NVSS (green crosses). The vertical dotted lines delimit the ring in the 2-dimensional Fourier space, where the SRT and the NVSS spatial scales overlap. The corresponding baselines range from $D=35$\,m to $D=64$\,m, that correspond to angular scales, $\lambda/D$, ranging respectively from 21\,arcmin to 11\,arcmin. The range of spatial scales contained in the images goes from the angular size of the SRT FoV ($\simeq 3\,{\rm deg}$), down to the resolution of the NVSS (45\,arcsec). Note that the Fourier amplitude profile of NVSS at short baselines ($<10$\,m) is comparable to that of SRT, although the NVSS contains no information on these scales due to the lack of short spacings. Indeed, this is a non-physical flux density caused by the convolution in Fourier space with the square window function transform determined by the FoV of the SRT image. However, this does not affect the result of the combination of the two Fourier transforms because we set the same FoV window function in the NVSS image as well. In fact, SRT does contain more large-scale flux density and actually helps recover extended structures from the Coma cluster, as will be shown below, but it is difficult to see the contribution directly on the amplitude profiles as this is small compared to the flux density of the numerous point sources in the field.
Finally, for proper combining, before merging we deconvolved the Fourier transforms of the SRT and NVSS images from the respective Gaussian beams. This causes the steep increase of the amplitude's spectra noticeable at high wavenumbers, which however will be compensated by the final re-convolution with the interferometer's synthesized beam. The SRT amplitude must be increased by a factor of 11\% to match the NVSS flux density. The upper panel, in the bottom-left plot of Fig.\,\ref{fig:srt_over_nvss}, shows the Fourier amplitude profiles after applying the scale factor to the SRT data. The insets show the consistency of the NVSS and SRT flux density in the ring.  In the  bottom-right plot of Fig.\,\ref{fig:srt_over_nvss} we present the merging of the Fourier transforms.  We used a linear weighting of the interferometer and single-dish amplitudes across the overlap ring, as represented in the lower panel by the shaded regions and the right-axis scale. The upper panel shows the final merged Fourier amplitude re-convolved to the 45\,arcsec angular resolution of the NVSS. The black contours in the top-right panel of Fig.\,\ref{fig:srt_over_nvss} represent the combination of the SRT and NVSS images after their merging in the Fourier space. Neither the radio halo nor the relic are detected at 45\,arcsec resolution in the SRT+NVSS image. The sensitivity of the NVSS is not
enough to detect these extended sources. The red tones and the gray contours show the residual contribution from the SRT data alone, obtained by subtracting the NVSS image from the SRT+NVSS image at 45\,arcsec resolution. It is clear that there is a significant  contribution to the large-scale flux density from the SRT to both the radio halo region  at the cluster center ($S_{\nu}\sim 400$\,mJy) and the peripheral relic region ($S_{\nu}\sim 200$\,mJy). However, in terms of surface brightness, these contributions are fainter than $\le 0.45$\,mJy/beam at 45\,arcsec resolution, which happens to be exactly the NVSS 1-$\sigma$ noise level.
Therefore, the NVSS is useful for providing a reliable calibration correction factor for the single-dish image, but cannot help detect the Coma radio halo and relic.

To obtain the deeper images necessary to reveal the halo and the relic, we combined the SRT image with pointed VLA observations  by keeping fixed the calibration correction derived from the NVSS. For the radio halo, we used the VLA pointed D-configuration archival observation at 1.49\,GHz  (project AG200) centered on 5C\,4.81. We present the Fourier combination of the VLA and SRT images in Fig.\,\ref{srt_over_vlaD_halo}. Note that the FoV is smaller than in Fig.\,\ref{fig:srt_over_nvss}. For this purpose, we produced an SRT image at a central frequency of exactly 1.49\,GHz and a bandwidth of 100\,MHz, to match the frequency coverage of the VLA observation. In the top-left panel, we show the VLA D-configuration image at 1.49\,GHz corrected for the primary beam attenuation. Before the combination, we tapered the SRT image exactly as the VLA image, basically by zeroing all pixels outside a circle of 1$^{\circ}$ in diameter centered in the VLA pointing.
We show the Fourier merging in the bottom panel. We kept fixed the scale factor of 11\% for the SRT flux density scale, as derived from the NVSS, and then we merged the Fourier amplitudes using the same linear weighting scheme as in Fig.\,\ref{fig:srt_over_nvss}. The SRT and VLA amplitudes match in the overlapping ring, but now the SRT provides a substantial contribution to angular scales larger than 20\,arcmin. We show the result in the top-right panel of Fig.\,\ref{srt_over_vlaD_halo}, where we present the combined VLA+SRT image at a resolution of 60\,arcsec. This image reveals a diffuse low-surface brightness halo at the cluster center. The halo has a surface brightness of about $\sim 0.5$ mJy/beam at an angular resolution of 1\,arcmin FWHM beam and 
 a linear size of about 500\,kpc (or 20\,arcmin in angular size). We discuss how the properties of the Coma radio halo compare to those of other halos from previous studies in
 Sect.\,\ref{profiles}. The  intensity is slightly above the $3\sigma_{I}$ level of the VLA D-array image but below the sensitivity of the NVSS, which explains why it was not detected in the image shown in Fig.\,\ref{fig:srt_over_nvss}. The extent of the radio halo detected in the SRT+VLA image at 1.49\,GHz is reassuringly similar to that observed at 1.4\,GHz by \citet{kim90} at a resolution of $71\arcsec\times 60\arcsec$ (see Fig.\,2 in that paper). They combined two interferometers, the DRAO Synthesis Telescope and the VLA, to recover the large scale emission, while we used the SRT single-dish antenna in combination with the VLA. The final results obtained with the two techniques are in excellent agreement.

We produced a combined SRT and VLA image for the radio relic 1253+273 as well. For this, we recovered from the VLA archive four deep D-configuration pointings at 1.46\,GHz (projects AA22 and AG200). We combined the four pointings using AIPS task LTESS. We show the resulting image in the top-left panel of Fig.\,\ref{srt_over_vlaD_relic}. We subtracted the strong point source Coma A from both the VLA mosaic and the SRT image, and we merged the Fourier amplitudes as shown in the bottom panel of Fig.\,\ref{srt_over_vlaD_relic}. Also in this case, we retained 11\% the scale factor of the SRT amplitude, derived from the NVSS. We present  the final SRT+VLA image of the radio relic at 1.46\,GHz and 60\,arcsec resolution in the top-right panel of Fig.\,\ref{srt_over_vlaD_relic}. Differently from what was observed for the central radio halo, in the case of the relic 1253+273 the addition of SRT data does not reveal the presence of new large scale emission. Given the morphology of the relic, the VLA does not lose flux density, as its short baselines are sufficient. The  Fourier amplitudes for the SRT and VLA data are comparable on the angular scales range from 20\,arcmin to 40\,arcmin, and consequently the shape of the radio relic is the same, apart from a slight increase in the transverse size. We observe some enhancement of the surface brightness only in the region in between the relic and the head-tail source associated to NGC\,4789. Consistent with the LOFAR images at 144\,MHz presented in \citet{bonafede21} and \citet{bonafede22}, the radio source's tails appear to be elongated towards the relic and in direct contact with it.

\begin{figure*}
    \centering
\includegraphics[width=0.6\textwidth,trim=120 105 120 10]{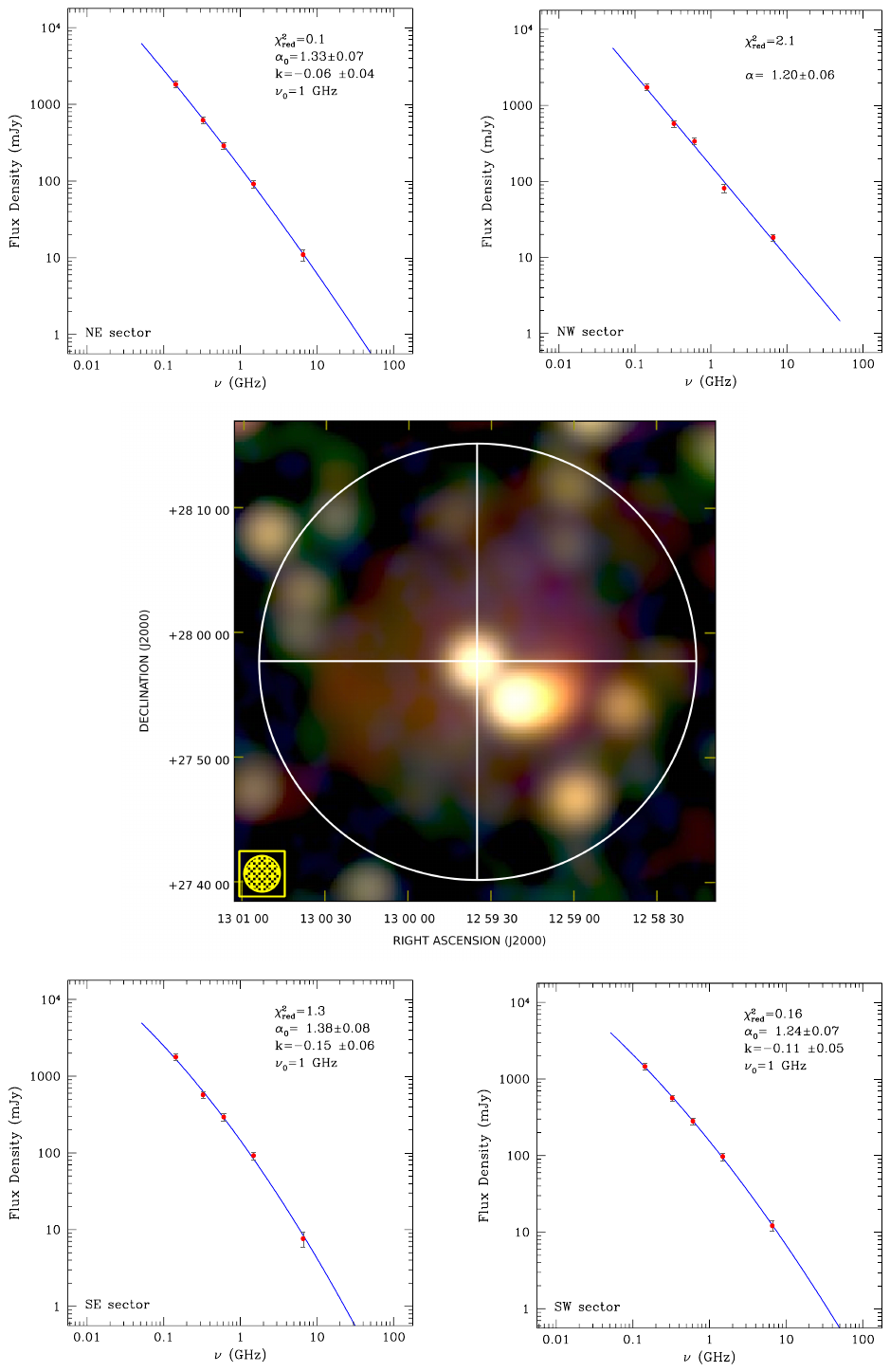}

    \caption{Tricolor radio halo image of the Coma radio halo obtained by superposing images at 0.608\,GHz (red), 1.49\,GHz (green), and 6.6\,GHz (blue) at a resolution of 2.9\,arcmin. The lateral panels show the halo spectra in the four sectors represented in the central picture, where we also included the LOFAR flux densities at 0.144\,GHz. The blue lines are the best fit of the modified power law model, except for the NW sector, where it represents the fit of a simple power law with no curvature.  We corrected the 6.6\,GHz flux density for the SZ decrement. The diameter of the circular region is the same we used to integrate the total flux density of the radio halo. Note that we have masked the regions occupied by discrete radio sources so that the radio spectra refer only to the diffuse emission.}
    \label{halo_local_spectrum}
\end{figure*}

\section{Analysis - Radio Halo}
\subsection{Radio halo flux density at 6.6\,GHz}
\label{radio_analysis_6.6}
We analyzed the SRT image at 6.6\,GHz to obtain an accurate measurement for the flux density of the radio halo. We integrated the total intensity over a circular aperture of 35\,arcmin in diameter, corresponding to about 1\,Mpc in linear size. The circumference delimits exactly the $3\sigma_{I}$ isophote. We included the contribution of both the discrete radio point sources (PS) and the diffuse emission. We used the accurate beam area of the SRT at 6.6\,GHz, $\Omega_{\rm beam}=9.93$\,arcmin$^2$ (see Appendix\,\ref{appendixD}), and we found a total (halo+PS) flux density of $S_{\rm tot}=194\pm5$ mJy over an area of $N_{\rm beam}=97$. 

We corrected the radio halo flux density for the thermal Sunyaev-Zel'dovich effect \citep[SZE;][]{sz72}. The thermal SZ effect causes a change in the apparent brightness of the cosmic microwave background (CMB) radiation towards the center of the galaxy cluster. At cm wavelengths, in the Rayleigh-Jeans approximation, the SZ-effect manifests as a decrement in the CMB temperature of $\Delta T_{\rm RJ}=-2y\cdot T_{\rm CMB}$. The Comptonization parameter, $y$, is given by the line-of-sight integral of the product of the thermal electron density, $n_e$, and temperature, $T_e$:
	\begin{equation}
	\label{y-parameter}
	y=\frac{\sigma_{T}}{m_{e}c^2}\int_{\rm LOS} n_e k_{B} T_e dl.
	\end{equation}

We calculated the intensity decrement at radio frequencies according to:
	\begin{equation}
	\label{SZ}
		\left(\frac{\Delta I_{\rm SZ}}{\rm mJy/beam}\right)= \frac{1}{340}\left(\frac{\nu}{\rm GHz}\right)^2 \left(\frac{\Delta T_{\rm RJ}}{\rm mK}\right) \left(\frac{\Omega_{\rm beam}}{\rm arcmin^2}\right),
	\end{equation}
see \citet{birkinshaw99} and \citet{basu16}. Using the image of the Comptonization parameter of the Coma cluster \citep{planck13}, we calculated an average value of $\langle y\rangle \simeq 4.3(\pm 1.0)\times 10^{-5}$ over the radio halo at 6.6\,GHz. By using $\Omega_{\rm beam}=9.93$\,arcmin$^2$ for the SRT, and  $T_{\rm CMB}=2725.5$\,mK for the CMB temperature \citep{fixsen09}, the expected intensity decrement is $\Delta I_{\rm SZ}=-0.30\pm 0.07$ mJy/beam. The corresponding correction for the SZ decrement at 6.6\,GHz in the circular aperture is $+29\pm 7$\,mJy. We cross-checked the residual base level of the SRT 6.6\,GHz image at the external radius of the radio halo using an elliptical annulus of one beam in width. The measured residual base level of the SRT image around the radio halo is $I_{\rm base}=-0.25\pm0.07$ mJy/beam. The average Comptonization parameter in the annulus is $\langle y\rangle=3.4\pm 0.9\times 10^{-5}$, and the expected SZ signal is $\Delta I_{\rm SZ}=-0.24\pm 0.06$ mJy/beam, which is fully consistent with the measured base level.

A significant fraction of the total flux density is due to the radio galaxies embedded in the radio halo, and their contributions must be carefully removed. In Fig.\,\ref{fig:halo_ps} we show a VLA C-configuration archival image at 1.49\,GHz (program AF196) in gray tones with the SRT radio contours at 6.6\,GHz overlaid. We identify ten radio sources with a 6.6\,GHz flux density greater than about 1\,mJy. We present the integrated spectra of these radio sources in the insets. The dots are the measurements derived from the literature, while the blue line is the best fit of a synchrotron model that is used to estimate the expected flux density at 6.6\,GHz (see Tab.\,\ref{flux_halo_ps}). The integrated radio spectra of 5C\,4.81 and 5C\,4.86 are well described by the Continuous Injection (CI) model. The CI model has three free parameters: the injection spectral index, $\alpha_{\rm inj}$, the break frequency, $\nu_{\rm b}$, and the flux density normalization \citep{murgia99}. The spectral index moderately steepens by +0.5 at frequencies above the spectral break, $\nu_{\rm b}$. We fitted the spectra of the  remaining point sources with a simple power law with index $\alpha$. Adding together the estimated flux densities of all the ten radio sources, we found $S_{\rm PS}=181\pm5$ mJy. 

\begin{table*}
	\centering
	\caption{Flux density measurements of the discrete radio sources embedded in the halo shown in Fig. \ref{fig:halo_ps}. The flux densities at 6.6\,GHz are estimates.}
	\label{flux_halo_ps}
	\begin{tabular}{lcccll}
		\hline
	Source	& Label &Frequency       & $S_{\nu}$        & Instrument          & Reference \\
		& &(MHz)           & (mJy)            &                     &            \\
		\hline
5C\,4.81	& A& 150           &  2633$\pm$81      & TGSS            &      \citet{intema17} \\ 
            &  &608           &  955$\pm$2.6     & WSRT            &   \citet{giovannini93}\\
		    &  &1490          &  415$\pm$12      & VLA             & VLA  project AF196 \\
		    &  &4860          &   114$\pm$3      & VLA             & VLA  project AF196 \\
            &  &6600          &   83$\pm$4       & \emph{Estimate} & this work\\
            &  &8440          &   63$\pm$2       & VLA             & VLA  project AF196\\
        	\hline
5C\,4.84	& B&  150           &  776$\pm$24      & TGSS & \citet{intema17} \\ 
            &  &608           &   369$\pm$1.3     & WSRT            &   \citet{giovannini93}\\
		    &  &1490          &   200$\pm$6      & VLA       & VLA  project AF196 \\
		    &  &4860          &   74$\pm$2       & VLA        & VLA  project AF196 \\
            &  &6600          &   59$\pm$3       & \emph{Estimate}    & this work\\
            &  &8440          &   50$\pm$2       & VLA              & VLA  project AF196\\
        	\hline
5C\,4.077	& C &150           &  72$\pm$4      & TGSS & \citet{intema17} \\ 
            &  &608           &  20$\pm$0.7     & WSRT            &   \citet{giovannini93}\\
		    &  &1400          &   11$\pm$0.7      & NVSS       & \citet{condon98} \\
            &  &6600          &   2.8$\pm0.5$       & \emph{Estimate}    & this work\\
            \hline
NGC\,4858	& D &150           &  19$\pm$3      & TGSS & \citet{intema17} \\ 
            &  &608           &  10$\pm$0.8     & WSRT            &   \citet{giovannini93}\\
		    &  &1490          &  7.7$\pm$0.5      & VLA       & VLA  project AF196 \\
            &  &6600          &   4.3$\pm1.0$       & \emph{Estimate}    & this work\\
	        \hline
12\,59\,20 +28\,05\,03	& E& 608           &  2.16$\pm$0.95     & WSRT            &   \citet{giovannini93}\\
		                &  &1490          &   2.28$\pm$0.32      & VLA       & VLA  project AF196 \\
		                &  &3000          &  2.35$\pm$0.24       & VLASS        & VLASS \\
                        &  &6600          &  2.4$\pm$0.5       & \emph{Estimate}    & this work\\
\hline
12\,59\,13 +27\,58\,37 & F &1490          &    0.70$\pm$0.16      & VLA       & VLA  project AF196 \\
		                &  &4860          &   0.86$\pm$0.05       & VLA        & VLA  project AF196 \\
                        &  &6600          &    0.90$\pm$0.09      & \emph{Estimate}    & this work\\
  		\hline
5C\,4.070	& G &150           &  241$\pm$9      & TGSS & \citet{intema17} \\ 
            &  &608           &   72$\pm$1     & WSRT            &   \citet{giovannini93}\\
		    &  &1490          &   30$\pm$1      & VLA       & VLA  project AF196 \\
            &  &6600          &   8.0$\pm$0.7       & \emph{Estimate}    & this work\\
	        \hline
5C\,4.074	& H &150           &  491$\pm$16      & TGSS & \citet{intema17} \\
            &  &608           &  153$\pm$1.7     & WSRT            &   \citet{giovannini93}\\
		    &  &1400          &   73$\pm$2      & NVSS       & \citet{condon98} \\
            &  &6600          &   19.7$\pm$1.7       & \emph{Estimate}    & this work\\
	        \hline
12\,59\,52 +27\,50\,01	& I &608           &  2.7$\pm$0.5     & WSRT            &   \citet{giovannini93}\\
		                &  &1490          &   1.8$\pm$0.3      & VLA       & VLA  project AF196 \\
		                &  &4860          &   0.54$\pm$0.05       & VLA        & VLA  project AF196 \\
                        &  &6600          &   0.43$\pm$0.05       & \emph{Estimate}    & this work\\
                  \hline
12\,59\,40 +27\,51\,23	& J &608           &  5.7$\pm$0.5     & WSRT            &   \citet{giovannini93}\\
		                &  &1490          &   2.3$\pm$0.2      & VLA       & VLA  project AF196 \\
		                & & 4860          &   0.54$\pm$0.05       & VLA        & VLA  project AF196 \\
                        &  &6600          &  0.39$\pm$0.05       & \emph{Estimate}    & this work\\
            \hline
	\end{tabular}
\end{table*}

We calculated the net radio halo flux density at 6.6\,GHz to be:
\begin{eqnarray}
S_{\rm halo} &=& S_{\rm tot}-I_{\rm SZ}\cdot N_{\rm beam}-S_{\rm PS}=\\
&=&194(\pm 5)+97\cdot[0.30(\pm0.07)]-181(\pm 5)= \nonumber\\
&=&42\pm 10 \,\rm mJy,  \nonumber
\label{haloflux}
\end{eqnarray}
where the final quoted error represents the 1$\sigma$ statistical uncertainty obtained by adding in quadrature the individual errors. We included in quadrature the systematic error on the total halo flux density of 3\%. This includes both the systematic uncertainty of the flux density calibrator and the systematic error of the SRT at 6.6\,GHz (see Appendix\,\ref{appendixB}). 

\begin{table*}
	\centering
	\caption{Flux density measurements of the radio halo in the Coma cluster. All measurements have the point sources subtracted. 
	Col. 1: Frequency [MHz]; Col. 2: Flux density [mJy]; Col. 3: Type of instrument: Int=image produced from interferometric observations, SD=
	image produced from single dish observations; Col. 4: Flux density reference.}
	\label{halo_measurements}
	\begin{tabular}{ccll}
		\hline
		Frequency       & $S_{\nu}$       & Instrument          & Reference \\
		(MHz)           & (mJy)           &                     &            \\
		\hline
		   30.9         & 49000$\pm$10000 &   Int - Clark Lake    & \citet{henning89} \\ 
		   34.5         & 60000$\pm$11000  &  Int - Gauribinadur  & \citet{sastry83} \\
		   43.0         & 51000$\pm$13000 &    Int - Clark Lake   & \citet{hanisch_erickson80} \\
		   73.8         & 17000$\pm$12000 &    Int - Clark Lake   & \citet{hanisch_erickson80} \\
		  139           & 10600$\pm$600   & Int - WSRT             & \citet{pizzo10}  \\ 
	      144           & 12200$\pm40$    & Int - LOFAR            & \cite{bonafede22}\\
	      151           &  7200$\pm$800   & Int - Cambridge Low-Frequency Synthesis Telescope  & \citet{cordey85} \\ 
		  326           &  3810$ \pm$30   & Int - WSRT       & \citet{venturi90} \\
		  342           &  5300$ \pm 100$ & Int -WSRT        &  \cite{bonafede22} \\
		  408           &  1500$\pm$400   & Int - Northern Cross  & \citet{ballarati81}\\
		  408           &  2000$\pm$200   & DRAO+Effelsberg         & \citet{kim90}; Flux density from \citet{giovannini93} \\
		  430           &  2550$\pm$280   &  SD - Arecibo          & \citet{hanisch80} \\
		  608         &  1200$\pm$300   &  Int- WSRT             & \citet{giovannini93} \\
		  610           & 2830$\pm$300     & SD   - GBT            & \citet{jaffe79} \\
		  610           & 1200$\pm$500     & Int - WSRT            & \citet{valentijn78} \\
		  1380          &  530$\pm$50      & Int -DRAO+VLA              & \citet{kim90}; Flux density from \citet{giovannini93} \\
		  1400          &  640$\pm$35      & SD - Effelsberg       & \citet{deiss97} \\
		  1400          &  520$\pm$180     & SD - Arecibo          & \citet{hanisch80} \\
		 1490          & 424$\pm$7         & Int+SD - VLA+SRT               & This work\\
		  2675          &  109$\pm$28      & SD - Effelsberg       & \citet{thierbach03}, SZ corrected  \\ 
		  4850          &   32$\pm$12     & SD - Effelsberg        & \citet{thierbach03}, SZ corrected\\
        6600          &   42$\pm$10    & SD  - SRT              & This work,  SZ corrected\\
  		\hline
	\end{tabular}
\end{table*}

\subsection{Radio halo flux density at 1.49\,GHz}
We derived the radio halo flux density at 1.49\,GHz by analyzing the combined SRT+VLA image at 60\,arcsec resolution. A mask image is constructed by identifying the discrete sources in the VLA C array image at 1.49\,GHz, where the extended emission is not detected. The radio halo at 1.49\,GHz fades gradually with radius, therefore we integrated the radio intensity over circular apertures of increasing diameter centered at RA$_{J2000}$: 12$^h$59$^m$35$^s$ and DEC$_{J2000}$: +27$^d$57$^m$50$^s$. We computed the flux density as the average intensity times the aperture area, measured in beam units. The halo flux density increases by increasing the aperture, up to a value of $S_{\rm halo}=424\pm 7$\,mJy at a radius of 650\,kpc, beyond that radius the flux
density profile stabilizes. The halo flux density is consistent with the residual contribution of the SRT to the extended emission after subtracting the NVSS, shown in the upper right panel of Fig.\,\ref{fig:srt_over_nvss}. Note, however, that at larger radii, the noise in the image is affected by the VLA primary beam correction and the integration of the flux density profile is not feasible (see Sect.\,\ref{profiles} for an alternative estimate  based on the extrapolation of the brightness radial profile). 

\subsection{Global spectrum of the radio halo}
\label{sec:halo_spectrum}
We compared our new measurements of the radio halo flux density at 1.49\,GHz and 6.6\,GHz with the data in the literature. We list the flux density measurements in Tab.\ref{halo_measurements}, along with the instruments used to collect the data. The quoted uncertainties refer to the 1$\sigma$ level. However, in the fit procedure, we include a systematic error of 10\% to account for the different flux density scales used in these works. By following the procedure used for the SRT measurement at 6.6\,GHz, we corrected the measurements at 2.675 and 4.85\,GHz for the SZ decrement, by adding respectively 1.9 and 5.9\,mJy to the original flux density of 107 and 26\,mJy reported by \citet{thierbach03}. At lower frequencies, the correction is negligible. We note that similar approaches to the one used in this work have been followed in the past to correct Effelsberg data for the SZE and to evaluate its impact on the interpretation of the spectral steepening observed in the Coma halo \citep{ensslin02,donnert10,brunetti13}.

We present the global radio spectrum of the Coma halo in Fig.\,\ref{halo_global_spectrum}. The spectrum is a collection of measurements taken over more than four decades with very different radio telescopes as well as different procedures to estimate fluxes, and this is reflected in the significant scatter of the data. 

We modeled the radio spectrum with a simple modified power law of the form:

\begin{equation}
\log_{10}(S_{\nu})=\log_{10}(S_0)-\alpha_0\cdot\log_{10}(\nu/\nu_0)+k\cdot[\log_{10}(\nu/\nu_0)]^2.
\label{halo_global_model}
\end{equation}

The model has three free parameters. The parameters $S_{0}$ and $\alpha_0$ represent the halo flux density and spectral index at the reference frequency of $\nu_0$, respectively. The parameter $k$ is the spectral curvature. We show the best fit model as a continuous line in Fig.\,\ref{halo_global_spectrum}, where we have used a reference frequency of $\nu_0=1$\,GHz. The best fit model yields a normalization of $S_0=743\pm 55$\,mJy. This is a robust estimate for the total radio halo flux density at 1\,GHz, given the large scatter of the individual measurements. The radio spectrum is steep. The best-fit spectral index at 1\,GHz is $\alpha_0=1.48\pm0.07$. However, the best-fit requires a significant curvature in the radio spectrum, with $k=-0.17\pm 0.07$. In this modified power law model, the spectral
index varies with frequency if the curvature $k \ne 0$.

We obtain the variation of the model spectral index with frequency, $\alpha_{\nu}$, by differentiating Eq.\,\ref{halo_global_model} with respect to $\log_{10}(\nu/\nu_0)$:

\begin{equation}
\alpha_{\nu}\equiv -\frac{d\log_{10}(S_{\nu})}{d\log_{10}(\nu/\nu_0)}=\alpha_0-2k\cdot \log_{10}(\nu/\nu_0).
\label{halo_global_model_spix}
\end{equation}

This logarithmic derivative can be viewed as the model spectral index calculated between an infinitesimally close pair of frequencies centered at frequency $\nu$. The curvature parameter $k$ can be viewed as the derivative of the spectral index with 
respect to log frequency:

\begin{equation}
k\equiv -\frac{1}{2}\cdot\frac{d\log_{10} \alpha_{\nu}}{d\log_{10}(\nu/\nu_0)}.
\label{halo_global_model_k}
\end{equation}

We represent the variation of the model spectral index with a dashed line in Fig.\,\ref{halo_global_spectrum}. We observe a progressive steepening of the radio halo spectrum, from a spectral index of 1.1 at 0.1\,GHz up to a spectral index of 1.8 at 10\,GHz. Indeed, we confirm the halo spectrum is curved, in agreement with previous results obtained with the Effelsberg 100-m telescope \citep{thierbach03}.

\begin{figure*}
    \centering
    \includegraphics[width=0.95\textwidth,trim=50 370 50 10]{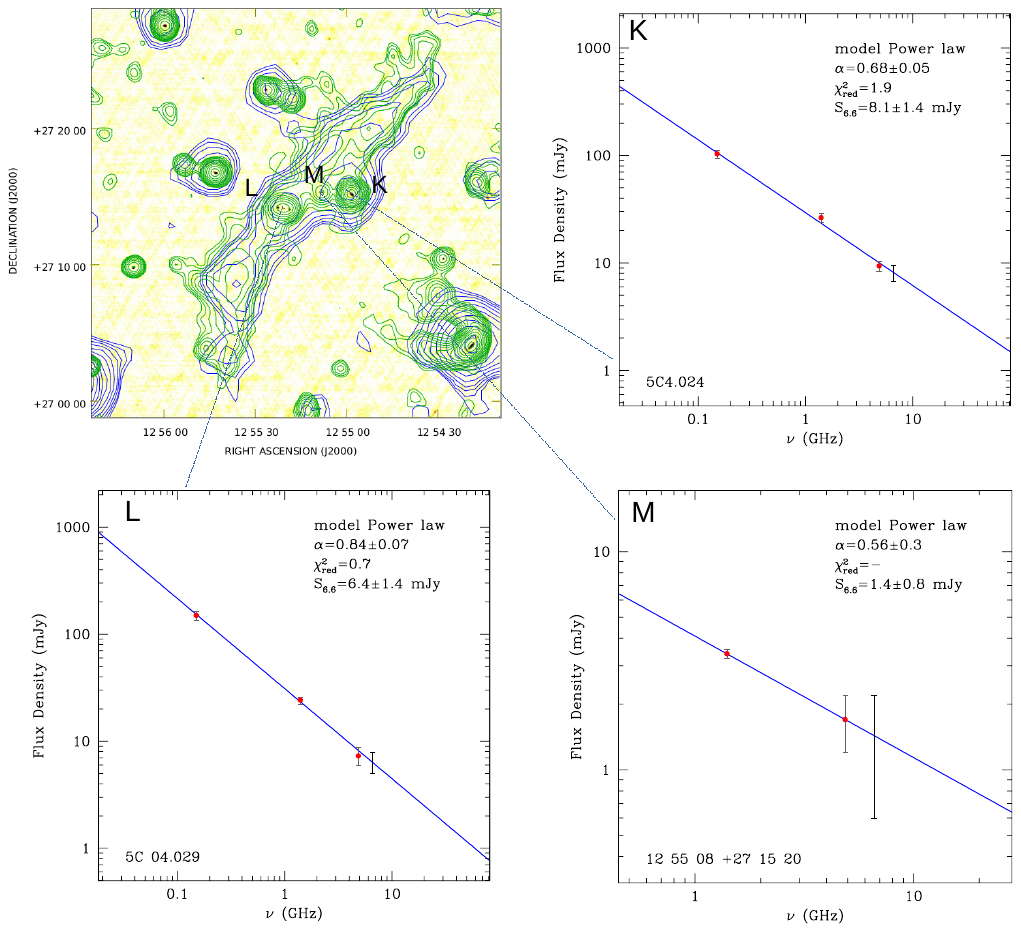}

    \caption{VLA FIRST image at 1.4 GHz and 6\,arcsec resolution of the relic 1253+273, with the SRT+VLA radio contours at 1.46\,GHz (green) and the SRT contours at 6.6\,GHz (blue) overlaid. The panels show the spectra of three discrete radio sources embedded in the radio relic. The bar shown in each spectrum marks the expected flux density at 6.6\,GHz. The blue line is the best fit of a power law model.}
    \label{fig:relic_ps}
\end{figure*}

\subsection{Radio halo spatially resolved spectra}

We extracted the spatially resolved spectra in four independent sectors of the radio halo. We used the LOFAR 144\,MHz image  \citep{bonafede22}, the WSRT 608\,MHz image \citep{giovannini93}, the WSRT 326\,MHz images \citep{venturi90}, and the SRT+VLA 1.49\,GHz and SRT 6.6\,GHz images from this work. We convolved the images to a common angular resolution of 2.9\,arcmin. We then extracted the spectra using the circular sectors shown in Fig.\,\ref{halo_local_spectrum}. The diameter of the circular area (35\,arcmin) is the same we used to integrate the total radio halo flux density at 6.6\,GHz.
We show the radio spectra of the four sectors in the lateral panels. We subtracted the contribution of point sources by masking the areas they occupy. For the calculation of the flux density, we assumed that these blanked area have a total intensity equal to the average in that sector. In practice, we multiplied the total intensity average by the number of beams contained in the entire sector. We also subtracted the flux densities at 6.6\,GHz of point sources labelled F for sector NW, and sources I and J for sectors SE. The contribution of these sources is relevant only at 6.6\,GHz, and thus we decided not to blank them in the images at 2.9\,arcmin resolution. We fit the local spectra of the NE, SE, and SW sectors, with the modified power law model in Eq.\ref{halo_global_model}. The local spectrum has in general a negative curvature, consistently with the integrated global spectrum. However, the NW sector is an exception. In the NW sector, the spectrum is best fit by a simple power law with no curvature. The simple power law fit still yields a better fit with a reduced $\chi^2=2.1$, while the curved power law has reduced $\chi^2=2.5$. We note that radio halo intensity at 6.6\,GHz is also, on average, higher in the NW sector. There may be a connection between the spectral and polarization properties observed in the NW sector. In fact, the polarized spot shown in Fig. 1 is located exactly within the NW sector, where no significant evidence of spectral curvature is observed. We do not have a definitive explanation for this coincidence. The power-law spectrum may suggest the possibility that the polarized emission is an external structure projected onto the halo, similar to what was found by \citet{rajpurohit21} for the case of MACS\,J0717.5+374. However, if existing, this elusive structure is probably diffuse and no easily distinguishable from the halo, as no obvious discrete sources associated with the polarized spot are found (see Appendix\,\ref{appendixC}). Another possibility could be found in the SRG/e-Rosita X-ray data presented by \citet{churazov21} showing the presence of a prominent secondary shock west of the core of the Coma cluster. The NW sector
is located in the downstream part of the shock. Therefore, the spectral and polarization properties we observe in this region could be related to the shock passage that possibly influenced both the magnetic field and the spectrum of the synchrotron emission.

We note that the spectral index $\alpha_0$, measured for the spectra of the four sectors, is flatter than the global one. However, while the global spectrum is a set of inhomogeneous measurements, the flux densities in the four sectors are obtained from images at the same resolution, in the same regions, and with the point sources subtracted.

\section{Analysis - Radio Relic}
\subsection{Radio relic flux density at 6.6\,GHz}
We measured the flux density of the radio relic at 6.6\,GHz by integrating the total intensity above the $3\sigma$ isophote. We found a total flux density of $S_{tot}=39.8\pm 1.2$ mJy, over an area of $N_{beam}=12.7$. We identified three point sources in the relic: 5C\,04.024, 5C\,04.029, and J125508+271520, see Fig.\,\ref{fig:relic_ps}. We list the flux densities of these sources in Tab.\,\ref{tab:flux_relic_ps}, and we show their radio spectra in the insets of Fig.\,\ref{fig:relic_ps}. We fitted the radio spectra using a power law, and we estimated the expected flux density at 6.6\,GHz. Considering all the three sources together, their contributions are $S_{\rm PS}=15.9\pm 2.1$\,mJy. 

We calculated that the SZ signal at the relic location is $\Delta I_{SZ}=-0.04\pm0.01$\,mJy/beam.
By correcting for the SZ negative signal and by subtracting the point sources contribution, we found for the net radio relic flux density at 6.6\,GHz:
\begin{eqnarray}
S_{\rm relic} &=& S_{\rm tot}-I_{\rm SZ}\cdot N_{\rm beam}-S_{\rm PS}=\\
&=&39.8(\pm 1.2)+12.7\cdot[0.04(\pm0.01)]-15.9(\pm 2.1)= \nonumber\\
&=&24.4\pm 2.4 \,\rm mJy,  \nonumber
\label{haloflux}
\end{eqnarray}

The quoted error refers to the statistical $1\sigma$ uncertainty, we neglected the systematic uncertainty because it is negligible given the relatively low flux density of the radio relic.

\subsection{Radio relic flux density at 1.46\,GHz}
We measured the radio relic flux density at 1.46\,GHz using the SRT+VLA image by integrating the total intensity down to the $3\sigma_{I}$ isophote. We found a total flux density of $S_{\rm tot}=230\pm2$ mJy, over an area of $N_{\rm beam}=121.7$. We subtracted from this value the point sources contribution of $S_{\rm PS}=49\pm 4$ mJy. 
We estimated a net relic flux density at 1.46\,GHz of $S_{\rm relic}=181\pm4$ mJy.
The relic flux density is consistent with the residual contribution of the SRT to the extended emission after subtracting the NVSS, shown in the upper right panel of Fig.\,\ref{fig:srt_over_nvss}.

\subsection{Global spectrum of the radio relic}
\label{sec:relic_spectrum}
\begin{figure}
    \centering
    \includegraphics[width=1.0\columnwidth,trim=25 200 75 125] {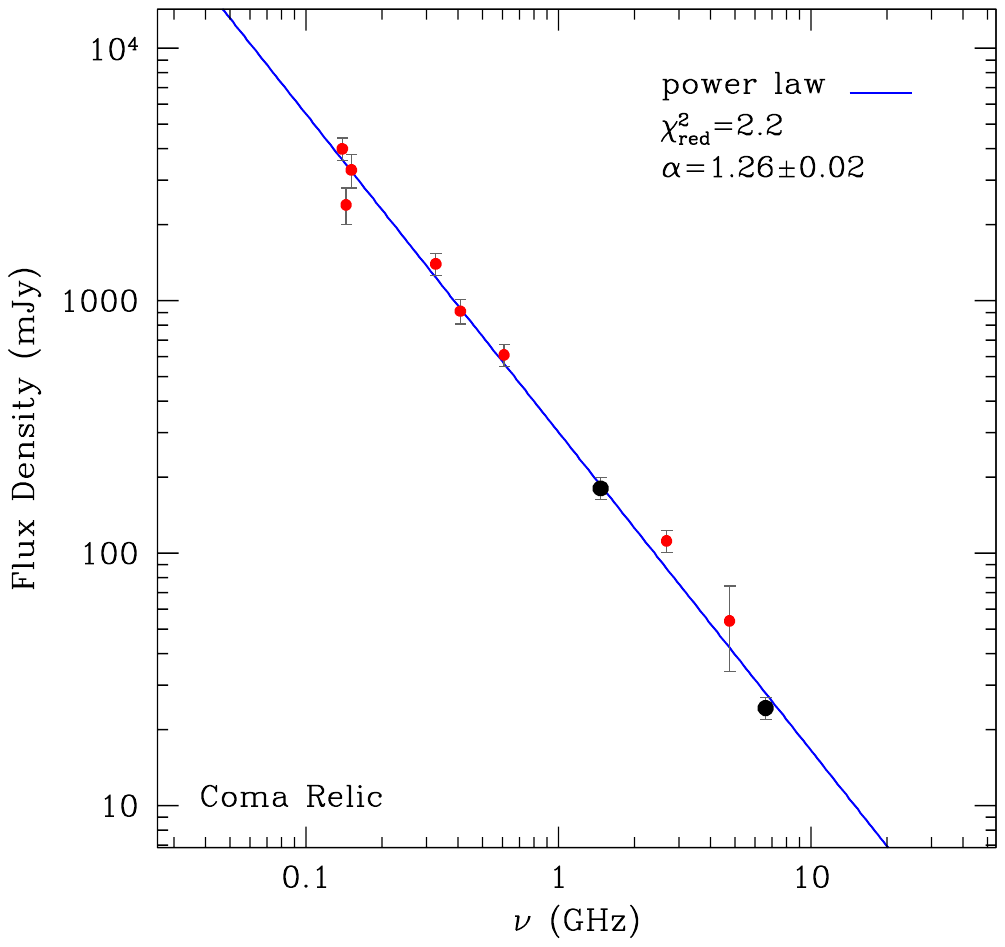}
    \caption{Flux density of the Coma relic as a function
of the frequency. Red dots are measurements from the literature, while black dots are the new 1.46\,GHz and 6.6\,MHz measurements presented in this work. The blue line represents the best fit of a power law model.
}
    \label{fig:relic_global_spectrum}
\end{figure}

We collected the flux density at different frequencies, available in the literature for the relic. These are listed in Tab.\,\ref{tab:flux_relic}, along with the measurements taken in this work. We present the global spectrum of the relic in Fig.\,\ref{fig:relic_global_spectrum}. The data set is made up by a collection of inhomogeneous flux density measurements taken over various decades with different instruments by different authors, and in the fitting procedure we considered for each data point a systematic error of 10\% error to account for the different calibration scales involved. Furthermore, we only included in Tab.\,\ref{tab:flux_relic} those measurements in which point sources have been subtracted from the flux density of the relic. 
The best fit of a power law model gives a global spectral index $\alpha=1.26\pm 0.02$ with a reduced $\chi^2_{\rm red.}=2.2$. We found a spectral index of the radio relic that is slightly steeper than the early value reported by \citet{thierbach03}, $\alpha=1.18\pm0.02$.

\begin{table*}
	\centering
	\caption{Flux density measurements of the discrete radio sources embedded in the relic, shown in Fig. \ref{fig:relic_ps}. The flux densities at 6.6\,GHz are estimates.}
	\label{tab:flux_relic_ps}
	\begin{tabular}{lcccll}
			\hline
	Source	& Label &Frequency       & $S_{\nu}$        & Instrument          & Reference \\
	&	&(MHz)           & (mJy)            &                     &            \\
		\hline
	5C\,04.024	& K &150           &  104$\pm$9      & TGSS            &\citet{intema17}     \\ 
		        &  &1400          &  26$\pm$3      & NVSS       & \citet{condon98} \\
		        &  &4860          &   9.4$\pm$1      & VLA             &  VLA project AG194 \\
	            &  &6600          &   8.1$\pm$1.4      & \emph{Estimate}    & this work\\
	            \hline
	5C\,04.029	& L& 150           &  150$\pm$15      & TGSS            &    \citet{intema17}   \\ 
		        &  &1400          &  24$\pm$2      & NVSS       &  \citet{condon98}\\
		        &  &4860          &   7.3$\pm$1.4      & VLA             & VLA project AG194 \\
	            &  &6600          &   6.4$\pm$1.4      & \emph{Estimate}    & this work\\
	            \hline
J12\,55\,08 +27\,15\,20    &  M&1400          &   3.4$\pm$0.2      & FIRST       &  \citet{becker95}\\
		        &  &4860          &   1.7$\pm$0.5      & VLA             &  VLA project AG194 \\
	            &  &6600          &   1.4$\pm$0.8      & \emph{Estimate}    & this work\\
	            \hline        
	
	\end{tabular}
\end{table*}

\begin{table*}
	\centering
	\caption{Flux density measurements of the relic shown in Fig. \ref{fig:relic_global_spectrum}. All measurements are with embedded point sources subtracted. Col. 1: Frequency [MHz]; Col. 2: Flux density [mJy]; Col. 3: Type of instrument: Int=image produced from interferometric observations, SD=
	image produced from single dish observations; Col. 4: Flux density reference.}
	\label{tab:flux_relic}
	\begin{tabular}{ccll}
		\hline
		Frequency       & $S_{\nu}$      & Instrument                   & Reference \\
		(MHz)           & (mJy)          &                              &           \\
		\hline
	      139           & 4000$\pm$200   & Int - WSRT                                    & \citet{pizzo10} \\ 
	    144            &2400$\pm 400$  & Int - LOFAR                         & \citet{bonafede22}\\
        151           & 3300$\pm$500   & Int - Cambridge Low-Freq. Synth.Tel. & \citet{cordey85}; Flux density from \citet{giovannini91} \\
	      326           & 1400$\pm$30    & Int - WSRT                                    & \citet{giovannini91} \\
		  408           & 910$\pm$100   & Int - Northern Cross                          & \citet{ballarati81}; Flux density from \citet{giovannini91}   \\
		  608           & 611$\pm$50     & Int - WSRT                                    & \citet{giovannini91} \\
 	      1465          & 181$\pm$4    & Int+SD  - VLA + SRT                                     & This work\\
		  2675          & 112$\pm$10     & SD - Effelsberg                               & \citet{thierbach03} \\ 
		  4750          &  54$\pm$20     & SD - Effelsberg                               & \citet{andernach84}   \\
          6600          & 24.4$\pm$2.4       & SD  - SRT                                     & This work, SZ corrected\\
  		\hline
	\end{tabular}
\end{table*}

\section{Discussion}

\subsection{The Coma radio halo in the $I_0-r_e$ plane}
\label{profiles}
We compared the basic morphological properties of the radio halo in the Coma cluster with those of other similar sources known in the literature. In particular, we used the SRT+VLA image at 1.49\,GHz to determine the central intensity and the $e$-folding radius of the diffuse emission.

\begin{figure*}
    \centering
    \includegraphics[width=1.0\textwidth,trim=90 370 60 10]{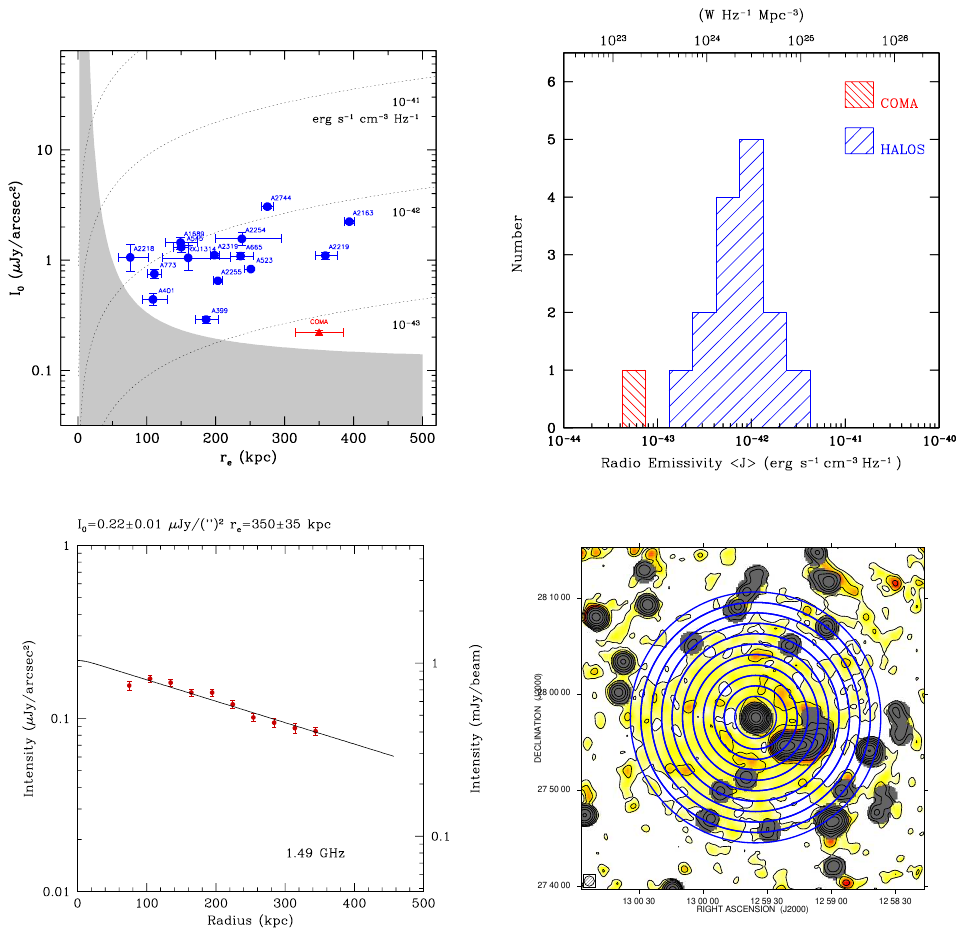}
    \caption{Top-left: $I_0-r_e$ plane at 1.4\,GHz for a sample of known radio halos, including the one in the Coma galaxy cluster. Radio emissivity is constant along the dotted lines. The gray area represents the undetectability region for a radio halo at redshift $z>0.1$ observed at a resolution of 60\,arcsec due to the confusion noise, see text. Top-right: distribution of radio emissivities for the sample of radio halos. Bottom-left: Intensity radial profile at 1.49\,GHz for the radio halo in the Coma cluster. Bottom-right: the set of circular annuli used to derive the radial profile. Regions containing discrete radio sources (gray areas) have been excluded.}
    \label{fig:radial_profiles}
\end{figure*}

In order to investigate the morphological properties of the diffuse radio emission in clusters of galaxies, \citet{murgia09} proposed to fit their azimuthally averaged intensity profile with an exponential of the form:
\begin{equation}
I(r)=I_0 e^{-r/r_e}.
\label{exponential}
\end{equation}
By fitting the observed intensity profile one obtains the central brightness, $I_0$, and the $e$-folding radius, $r_e$, from which the average radio halo emissivity can be calculated as:

\begin{equation}
\langle J \rangle \simeq 7.7\times 10^{-41} (1+z)^{3+\alpha}\cdot\frac{I_0}{r_e}\quad\rm(erg\,s^{-1} cm^{-3}Hz^{-1}),
\label{emissivity}
\end{equation}
or, in more convenient units:
\begin{equation}
\langle J \rangle \simeq 2.3\times 10^{26} (1+z)^{3+\alpha}\cdot\frac{I_0}{r_e}\quad\rm(W/Hz/Mpc^{3}),
\label{emissivity2}
\end{equation}
where $I_0$ and $r_e$ are respectively measured in $\mu$Jy/arcsec$^2$ and kpc. The term $(1+z)^{3+\alpha}$ takes into account of both the $k$-correction and cosmological dimming of the surface brightness with redshift.

In the bottom-left panel of Fig.\,\ref{fig:radial_profiles} we show the modeling of the azimuthally averaged radial profile of the Coma radio halo at 1.49\,GHz. We obtained the profile using the circular annuli shown in the bottom-right panel of Fig.\,\ref{fig:radial_profiles}. We masked all the point sources, implicitly assuming for the blanked regions the same halo average brightness as in the rest of the annulus. By a fit of the exponential law, we obtained $I_0=0.22\pm0.01$ $\mu$Jy/arcsec$^2$ and $r_e=350\pm 50$\,kpc. The flux density obtained by extrapolating the surface brightness profile up to 3$r_e$ is of $S_{3r_e}=650\pm 95$\,mJy.

In the top-left panel of Fig.\,\ref{fig:radial_profiles} we show the Coma cluster in the  $I_0-r_e$ plane at 1.4\,GHz along with a sample of 15 known radio halos presented in \citet{murgia09}, \citet{murgia10}, and \citet{vacca11}. The radial profiles of all these radio halos have been analyzed consistently, and we found that in general radio halos can have quite different length-scales, but their emissivities are remarkably similar from one halo to the other. However, the Coma radio halo is different in that it appears to be much fainter of the other halos of comparable linear size.

 In the top-right panel of Fig.\,\ref{fig:radial_profiles} we show that the distribution of radio halo emissivities peaks at $\langle J \rangle \simeq 10^{-42}$ erg\,s$^{-1}$cm$^{-3}$Hz$^{-1}$ (or equivalently $\langle J \rangle \simeq 3\times 10^{24}$ W\,Hz$^{-1}$Mpc$^{-3}$). The Coma radio halo, with an emissivity of $\langle J \rangle= 5.3\times 10^{-44}$ erg\,s$^{-1}$cm$^{-3}$Hz$^{-1}$ (or $1.6\times 10^{23}$ W\,Hz$^{-1}$Mpc$^{-3}$), is the radio halo with the lowest emissivity observed so far.
 
Indeed, the position of the Coma radio halo in the $I_0-r_e$ plane is close to the shaded area that indicates the region of parameter space where a radio halo at redshift $z>0.1$ would be undetectable because of the confusion limit if observed at 60\,arcsec resolution. We traced this region by considering that a radio halo would not be detectable if its central brightness is fainter than the minimum intensity $I_0<I_{min}(z)$. To derive this limiting brightness,
we impose that the radius at which the intensity profile in  Eq.\,\ref{exponential} intercepts the $3\sigma_{c}$ noise floor is equal to the FWHM beam:

\begin{equation}
I_{min}(z) = 3\sigma_{c}\cdot \exp{\left[\frac{\theta_{\rm beam}\arcsec}{r_{e}(z)\arcsec}\right]},
\label{I0min}
\end{equation}
where $\theta_{\rm beam}$ is the FHWM beam and $r_e(z)\arcsec$ is the (redshift dependent) angular size of the e-folding radius in arcseconds, while for the confusion noise we adopted the formula by \cite{condon02}:
\begin{equation}
\sigma_c=0.05\cdot (\nu/\rm GHz)^{-0.8}(\theta_{\rm beam}/60\arcsec)^2 \quad \rm \mu Jy/arcsec^2
\label{haloconfusion}
\end{equation}

In practice, we require that, to be detectable as such, a radio halo must have a minimum angular diameter, measured using the $3\sigma_c$ isophote, of at least two times the FWHM beam. 

\begin{figure}
    \centering
    \includegraphics[width=0.45\textwidth,trim=50 170 100 200]{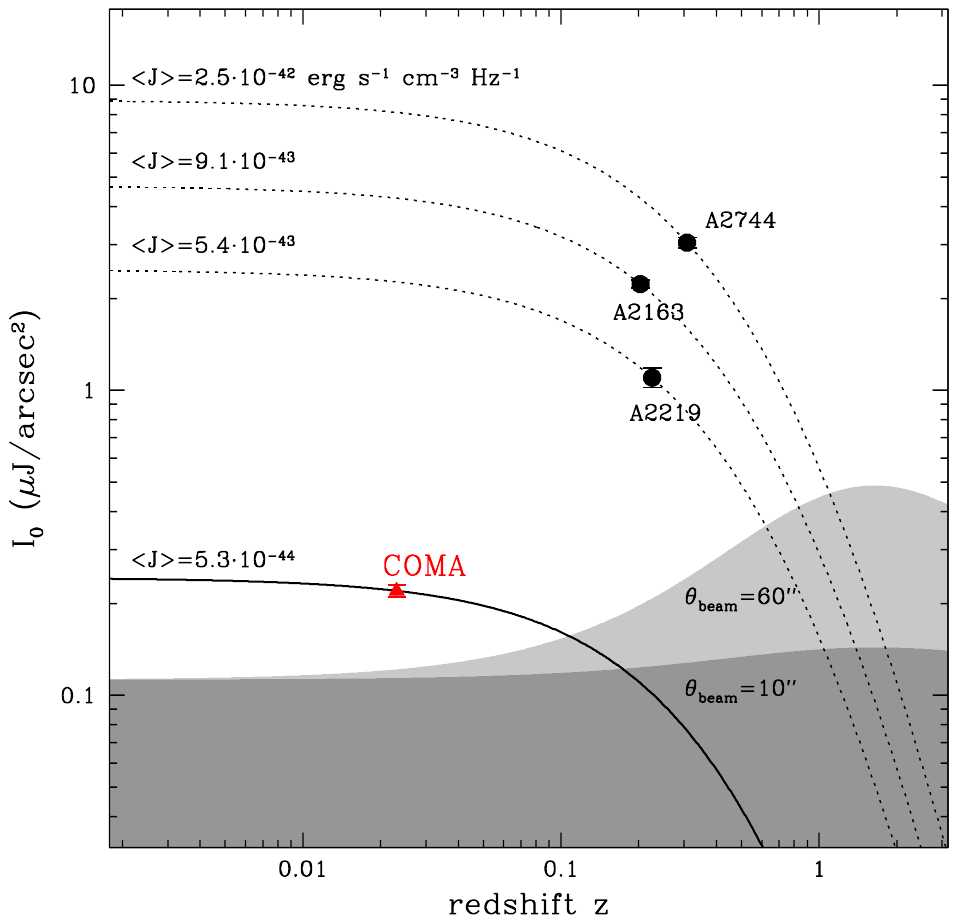}
    \caption{The Coma radio halo in the I$_0$-redshift plane in comparison with much brighter halos of similar e-folding radius. The lines represent the dimming of the radio halo surface brightness as a function of redshift for a constant emissivity. The shaded areas indicate the minimum intensity below which the radio halo would be undetectable at 1.4\,GHz because of the confusion limit. Light and dark gray areas correspond to angular resolutions of 60 and 10\,arcsec, respectively (see text).}
    \label{fig:I0-redshift}
\end{figure}

In Fig.\ref{fig:I0-redshift}, we show the Coma radio halo in the I$_0$-redshift plane in comparison with the much brighter halos, of similar size, hosted in the galaxy clusters A2219, A2163, and A2744. The average $r_e$ of these radio halos is about 340\,kpc, that is comparable to that of the Coma radio halo. The lines represent the dimming of the radio halo brightness as a function of redshift for a constant emissivity. We estimate that a Coma-like radio halo could be detectable only up to maximum redshift $z\lesssim 0.1$, beyond this distance it would be hardly observable at 1.4\,GHz and 60\,arcsec resolution. This region of parameter space is represented by the light-gray area. The interferometers of the new
generation (such as MeerKAT, ngVLA and SKAO) may allow for confusion limited images of radio halos at even higher angular resolution.
For example, In Fig.\,\ref{fig:I0-redshift} we show in dark-gray the predicted undetectable region for a radio halo observed at 1.4\,GHz and 10\,arcsec resolution. In this case, it would be possible to observe a Coma-like radio halo up to a redshift $z\lesssim 0.2$.

The Coma radio halo could be representative of a class of intrinsically large radio halos with low surface brightness that are visible only at very low redshifts. We speculate that the gap, in the $I_0-r_e$ plane shown in the top panel of Fig.\,\ref{fig:radial_profiles}, between the Coma and the other radio halos of comparable size could be due to this observational bias.

We compare our estimate for the halo central brightness of $I_0=0.22\pm0.01$\,$\mu$Jy/arcsec$^2$ at 1.49\,GHz, with the value of $I_0=5.42\pm0.04$\,$\mu$Jy/arcsec$^2$ at 0.144\,GHz, estimated using LOFAR by \citet{bonafede22}. The two values are indeed in excellent agreement if one considers for the radio halo a spectral index of $\alpha=1.4$ between 0.144 and 1.49\,GHz. This is the same spectral index value that is deduced using the model fit shown in Fig.\,\ref{halo_global_spectrum}. The e-folding radius at 1.49\,GHz, $r_e=350\pm35$\,kpc, is slightly larger than that at 0.144\,GHz,  $r_e=310\pm1$\,kpc. 

\subsection{The Coma radio relic shock wave}

Radio relics are associated with cosmological shock waves produced by accretion and merging phenomena in galaxy clusters \citep{ensslin98}. In the context of the Diffusive Shock Acceleration theory \citep{drury83}, the shock-wave compression ratio, $R$, is related to the \emph{stationary} radio spectral index by:
\begin{equation}
R=\frac{\alpha+1}{\alpha-1/2},
\label{copression_ratio}
\end{equation}
where $\alpha=\alpha_{\rm inj}+0.5$, and the injection spectral index is related to $\delta$, the index of the energy distribution of accelerated particles in the shock, by $\alpha_{\rm inj}=(\delta-1)/2$. The stationary spectral index refers to the asymptotic power law regime that is reached when the number of particles that leave a given energy bin because of the radiative losses is equally replaced by the newly accelerated particles. If the integrated radio spectrum of the Coma relic represents the steady state, then the observed high-frequency spectral index $\alpha=1.26\pm 0.02$ 
implies a compression factor $R=3.0\pm 0.1$. From the shock theory, see e.g. \citet{hoeft07}, we can relate the compression factor, $R$, to the upstream Mach number defined as:
\begin{equation}
M=\frac{v_u}{c_u},
\label{mach}
\end{equation}
where $v_{u}$ and $c_u$ are respectively the upstream gas velocity and sound speed.
By assuming a polytropic index $\gamma_{\rm gas}=5/3$, we have:
\begin{equation}
M^2=\frac{\alpha+1}{\alpha-1},
\label{mach_alpha}
\end{equation}
which gives $M=2.9\pm 0.1$. The relic spectral index is in fact steeper than the canonical value of $\alpha=1$ that would be expected for hypersonic shock waves, $M\gtrsim 10$.

The Mach number we derived from the radio spectrum is marginally consistent, within the uncertainties, with the values deduced from the analysis of the SZE caused by the pressure discontinuity in proximity of the relic: $M=2.9^{+0.8}_{-0.6}$ \citep{erler15} and $M=2.2\pm 0.3$ \citep{basu16}. However, as for many other relics in the literature, for the relic in Coma we also found that the radio Mach number is larger than values inferred from X-ray temperature discontinuity based on XMM data $M=1.9_{-0.40}^{+0.16}$ \citep{ogrean13}, Suzaku data $M=2.2 \pm 0.5$ \citep{akamatsu13}, and SRG/e-Rosita $M=1.9\pm0.2$ \citep{churazov23}.
Note that \citet{churazov23} derived a radio Mach number of $M=3.5$ from a spectral index  $\alpha=1.18\pm0.02$ \citep{thierbach03}. Based on the new radio data presented in this work, we find the integrated spectrum is better described by a steeper spectral index $\alpha=1.26\pm0.02$, which results in a radio Mach number slightly closer to the SZE and X-rays estimates.

From the mass conservation at the shock front, the downstream gas velocity is $v_d=v_u/R$. If the ICM behaves like an ideal polytropic gas, $c_{u}^2=\gamma_{\rm gas}k_BT_u/(\mu m_p)$, and we consider a mean molecular weight $\mu=0.6$, then $c_u\simeq 516\,(T_u/{\rm keV})^{1/2}$\,km/s. By considering a temperature of $T_u=1.54_{0.39}^{+0.45}$\,keV \citep{akamatsu13}, the upstream ICM sound speed is $c_u=640\pm 88$\,km/s. A radio Mach number of $M=2.9\pm0.1$ implies that the upstream and downstream gas velocities are respectively of $v_u=1860\pm 260$\,km/s and $v_d=620\pm110$\,km/s.

Following the reasoning of \citet{Markevitch05}, we relate the width of the synchrotron-emitting region, $D$, with the velocity of the down-stream flow that carries the relativistic electrons away from the shock. 
The relic transverse size change with frequency. In line with what is usually observed, the width is significantly larger
at low-frequencies. In the LOFAR image at 144 MHz we measure a width $D=220 \pm 30$ kpc from the 3-sigma isophote. In the SRT 6.6 GHz image the width is 80 kpc. In the analysis we decided to adopt the width measured 
in the LOFAR image at 144\,MHz because this is the one less subject to the effect
of the radiative losses of the synchrotron electrons.
The confinement time of the radio electrons within this thickness from the shock front is estimated as $t_{\rm dyn}=D/v_{d}\simeq 350\pm110$\,Myr, where $v_d$ is the downstream velocity.

\begin{figure}
    \centering
    \includegraphics[width=0.35\textwidth,trim=120 170 100 200]{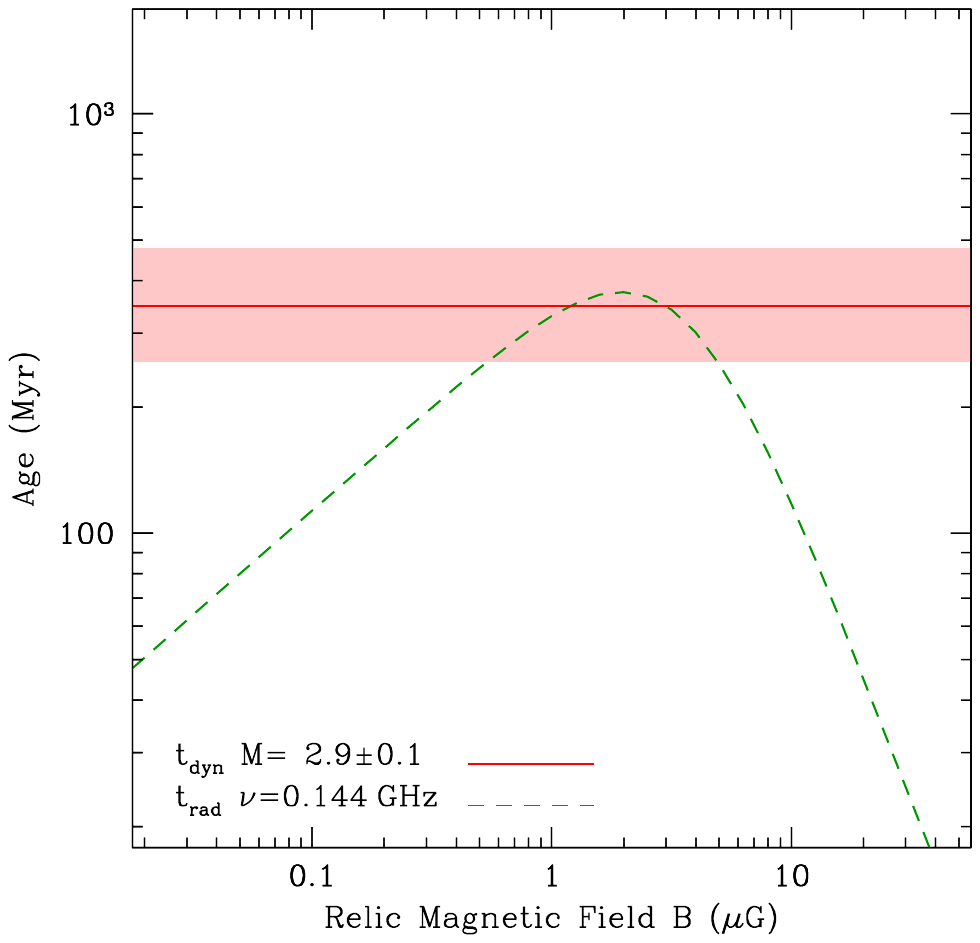}
    \caption{Confinement time ($t_{\rm dyn}$, red color) and radiative age ($t_{\rm rad}$, green color and dashed line) of the electrons in the relic as a function of the magnetic field strength. The confinement time is calculated from a relic width of $D=220\pm30$\,kpc by considering a Mach number $M=2.9\pm 0.1$ and an upstream sound speed $c_u = 640\pm88$\,km/s. The profile of the radiative time is calculated for a spectral break $\nu_b=144$\,MHz.}
    \label{fig:relic_age}
\end{figure}

According to \citet{ensslin98}, we then compare confinement time of the radio electrons in the relic to the radiative age due to 
synchrotron and inverse Compton energy losses:
\begin{equation}
t_{\rm rad}=1590\frac{B_{\mu\rm G}^{0.5}}{(B_{\mu G}^2+B_{\rm IC}^2)[(1+z)\nu_b/\rm GHz]^{0.5}}\quad \rm Myr,
\label{synage}
\end{equation}

where $\nu_b$ is the break frequency in GHz, $B$ is the relic magnetic field in $\mu$G, and $B_{\rm IC}=3.25(1+z)^2$\,$\mu$G is the inverse Compton equivalent magnetic field whose energy density is equal to that of the CMB photons. 
In Fig.\,\ref{fig:relic_age}, we show the confinement time $t_{\rm dyn}$ and the synchrotron age, $t_{\rm rad}$, as a function of the magnetic field strength, $B$. The radiative age is computed from Eq.\,\ref{synage} using a break frequency of $\nu_{b}=144$\,MHz, meaning that due to the radiative losses the spectral break shifted to the low end of the observed frequency range. This is the condition necessary to observe an integrated spectrum consistent with the stationary spectral index $\alpha=\alpha_{\rm inj}+0.5$. We note that $t_{\rm rad}$ is generally lower than $t_{\rm dyn}$, except for $B$ in the range from 0.5 to 5$\,\mu$G, where the two time-scales are comparable within the uncertainties. The closest match between the two time-scales is obtained around $B\simeq 2\,\mu$G. This is the magnetic field strength ($B=B_{\rm IC}/\sqrt{3}$) that maximizes the radiative age of the synchrotron electrons, i.e. $t_{\rm rad}\simeq 380 \pm {70}$\,Myr. Note that \citet{Bonafede13} inferred from Faraday rotation data that the magnetic field in the relic region is $\sim 2\,\mu$G.

A simple estimate of the energy dissipated in the relic can be obtained  by
considering the change in the kinetic energy flux across the shock:
\begin{equation}
    \Delta F_{\rm KE}=\frac{1}{2}\rho_{u} v_u^3\left(1-\frac{1}{R^2}\right)~~ \rm erg\,s^{-1}cm^{-2},
\end{equation}
 where $\rho_u$ is the upstream gas mass density \citep{finoguenov10}. Taking for the part of the shock-front directly associated with the radio relic a perpendicular circular area of 0.5\,Mpc${^2}$, the rate of kinetic energy dissipation is estimated as
 \begin{equation}
    \frac{d E_{\rm KE}}{dt}\simeq 1.3\times 10^{45}~~ \rm erg\,s^{-1}
    \label{kine}
\end{equation}
where we take $\rho_u=n_e m_p$, with $m_p$ the proton mass and $n_e=5.6\times 10^{-5}$\,cm$^{-3}$ is the thermal electrons density at the radial distance of the Coma relic from the cluster centre \citep{churazov21}.

The total non-thermal luminosity, $L_{\rm NT}=L_{\rm radio}+L_{\rm IC}$, of the relativistic electron
radiating by synchrotron and inverse Compton processes in the observed radio band (0.144 - 10\,GHz) is:
 \begin{equation}
L_{\rm NT}= 1.5\times 10^{40}\left[{1+\left(\frac{B_{IC}}{B_{\mu G}}\right)^2} \right]~~\rm erg\,s^{-1}
\label{Lnth}
\end{equation}

If this luminosity is sustained by the kinetic energy dissipated in the shock, then the efficiency of the electron acceleration is given by the ratio
 \begin{equation}
\epsilon\equiv \frac{L_{\rm NT}}{\frac{d E_{\rm KE}}{dt}}.
\end{equation}
By inserting  Eq.\,\ref{kine} and Eq.\,\ref{Lnth}, we get for the Coma relic:
\begin{equation}
\epsilon\simeq1.1\times 10^{-5} \left[{1+\left(\frac{B_{IC}}{B_{\mu G}}\right)^2} \right].
 \label{epsilon}
 \end{equation}

Indeed, considering the new value for the radio Mach number of $M=2.9\pm 0.1$, we find that a relatively small acceleration efficiency ($\epsilon\lesssim 1\%$) is required to match the radio luminosity observed for the Coma relic if  $B>0.1\,\mu$G. A similar result to what we find was obtained by \citet{finoguenov10} for the A3376 relic.
However, the required shock acceleration efficiency increases if the relic's power law spectrum extends to frequencies much lower than those observed so far. In particular, for others well studied radio relics \citet{botteon20} estimated that significantly higher values of acceleration efficiency are needed to explain the observed radio luminosity if the electrons are accelerated 
directly from the thermal pool. An alternative scenario often considered in the literature is the one in which the shock wave has
accelerated a population of electrons that are already at relativistic
energies. In the case of the Coma relic, there is a convincing evidence that plasma injected into the ICM from the head-tail source associated with NGC\,4789 could provide the seed relativistic electrons that the low Mach number shock wave would reaccelerate, see the discussion in \citet{bonafede22}. 

\section{Conclusions}
In this paper, we present deep total intensity and polarization observations of the Coma cluster of galaxies at 1.4 and 6.6\,GHz performed with the SRT. Our results are summarized as follows:

\begin{itemize}

\item{By combining the single-dish 1.4\,GHz data with archival VLA observations we obtain new images of the central radio halo and of the peripheral radio relic where we properly recover the flux density from the large scale structures.}

\item{In the halo, we find a localized spot of polarized signal at 6.6\,GHz, with fractional polarization of about 45\%, and a hint of polarized emission that extends along the north-east side of the diffuse emission.}

\item{The radio relic at 6.6\,GHz is highly polarized, up to 55\%, and the projected direction of the $\vec{E}$-field is perpendicular to the relic ridge line, as usually found for these sources.}

\item{The integrated halo spectrum shows evidence of spectral curvature, with a frequency spectral index of $\alpha=1.48\pm0.07$ at a reference frequency of 1\,GHz. However, because of the spectral curvature, the halo spectral index increases from $\alpha\simeq 1.1$, at 0.1\,GHz, up to $\alpha\simeq 1.8$, at 10\,GHz.}

\item{ We modeled the integrated spectrum of the radio relic using a  power law, and we find for the stationary spectral index $\alpha=1.26\pm 0.02$. From the shock theory, we inferred a radio Mach number $M=2.9\pm0.1$, which is higher than the estimates based on the SZE and the X-rays present in the literature. The radiative lifetime of the electrons emitting at low-frequency is comparable with the confinement time in the relic if the magnetic field in the emitting region is of the order of $B\simeq 2\,\mu$G.}

\item{We calculated that in the case of the Coma relic, the non-thermal luminosity of the relativistic electrons radiating in the radio band can be supported by a few percent of the kinetic energy dissipated in the shock wave if the magnetic field strength is $B> 0.1\,\mu$G.}

\item{We compared the Coma radio halo surface  brightness profile at 1.4\,GHz (central brightness, $I_0$, and e-folding radius, $r_e$) with the same properties of the other halos. We find that the Coma radio halo is peculiar in that while its radius is comparable to the intrinsically larger halos, its central brightness is one order of magnitude fainter. This means that the Coma radio halo has one of the lowest emissivities at 1.4\,GHz observed so far:  $\langle J \rangle \simeq 5.3\times 10^{-44}$ erg\,s$^{-1}$cm$^{-3}$Hz$^{-1}$, or equivalently $\langle J \rangle \simeq 1.6\times 10^{23}$ W\,Hz$^{-1}$Mpc$^{-3}$.}

\item{We calculated that a diffuse radio source with the properties of the Coma radio halo would be undetectable, at 1.4\,GHz and 60\,arcsec resolution, beyond a redshift $z\lesssim 0.1$ due to the confusion limit and the cosmological dimming of the surface brightness. At 10\,arcsec resolution, the redshift limit is $z\lesssim 0.2$.}
\end{itemize}

\section*{Data availability}
The data underlying this article will be shared on reasonable request to the corresponding author.

\section*{Acknowledgements}
We thank the anonymous referee for the useful comments and suggestions.
The Sardinia Radio Telescope is funded by the Ministry of Education, University and Research (MIUR), Italian Space Agency (ASI), the Autonomous Region of Sardinia (RAS) and INAF itself and is operated as a National Facility by the National Institute for Astrophysics (INAF). The Enhancement of the Sardinia Radio Telescope (SRT) for the study of the Universe at high radio frequencies is financially supported by the National Operative Program (Programma Operativo Nazionale - PON) of the Italian Ministry of University and Research "Research and Innovation 2014-2020", Notice D.D. 424 of 28/02/2018 for the granting of funding aimed at strengthening research infrastructures, in implementation of the Action II.1 – Project Proposals PIR01\_00010 and CIR01\_00010. The development of the SARDARA back-end has been funded by the Autonomous Region of Sardinia (RAS) using resources from the Regional Law 7/2007 "Promotion of the scientific research and technological innovation in Sardinia" in the context of the research project CRP-49231 (year 2011, PI Possenti): "High resolution sampling of the Universe in the radio band: an unprecedented instrument to understand the fundamental laws of the nature". VV and MM acknowledge support
from INAF mainstream project “Galaxy Clusters Science with LOFAR” 1.05.01.86.05. MB acknowledges support from the agreement ASI-INAF n. 2017-14-H.O and from the PRIN MIUR 2017PH3WAT “Blackout”. The National Radio Astronomy Observatory is a facility of the National Science Foundation operated under cooperative agreement by Associated Universities, Inc. This research has made use of the NASA/IPAC Extragalactic Database (NED), which is operated by the Jet Propulsion Laboratory, California Institute of Technology,
under contract with the National Aeronautics and Space Administration. Basic research in Radio Astronomy at the U.S.\ Naval Research Laboratory is supported by 6.1 Base funding.






\appendix
\section{Imaging stacking}
\label{appendixA}
\begin{figure*}
    \centering
    \includegraphics[width=0.75\textwidth,trim=25 425 25 15]{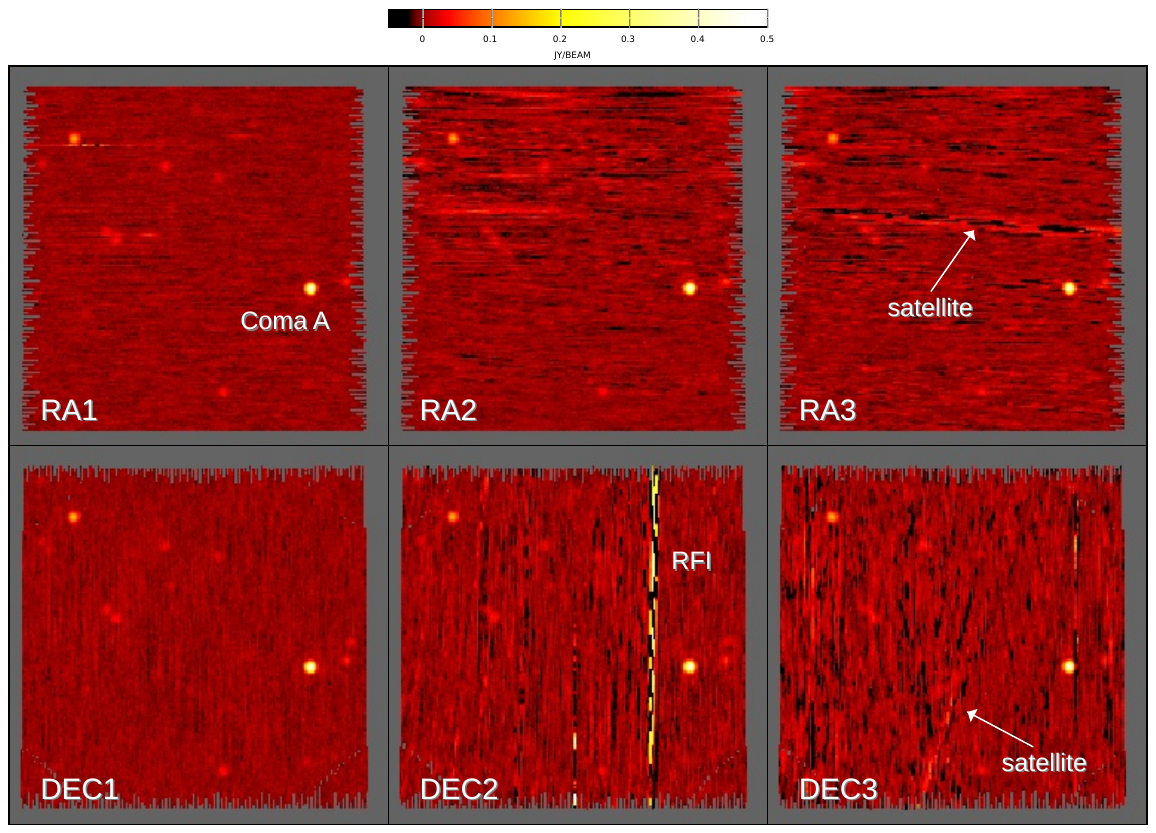}

    \caption{Spectral average of polarization $R$ for the C-band scans acquired on 06 Feb 16. Top panels: scans along the RA. Bottom panels: scans along DEC. RFI from satellites appear as oblique streaks while the RFI aligned with the scan direction, like in the DEC2 scan, originating most probably from ground sources.}
    \label{fig:radec}
\end{figure*}

\begin{figure*}
    \centering
    \includegraphics[width=0.75\textwidth,trim=25 425 25 15]{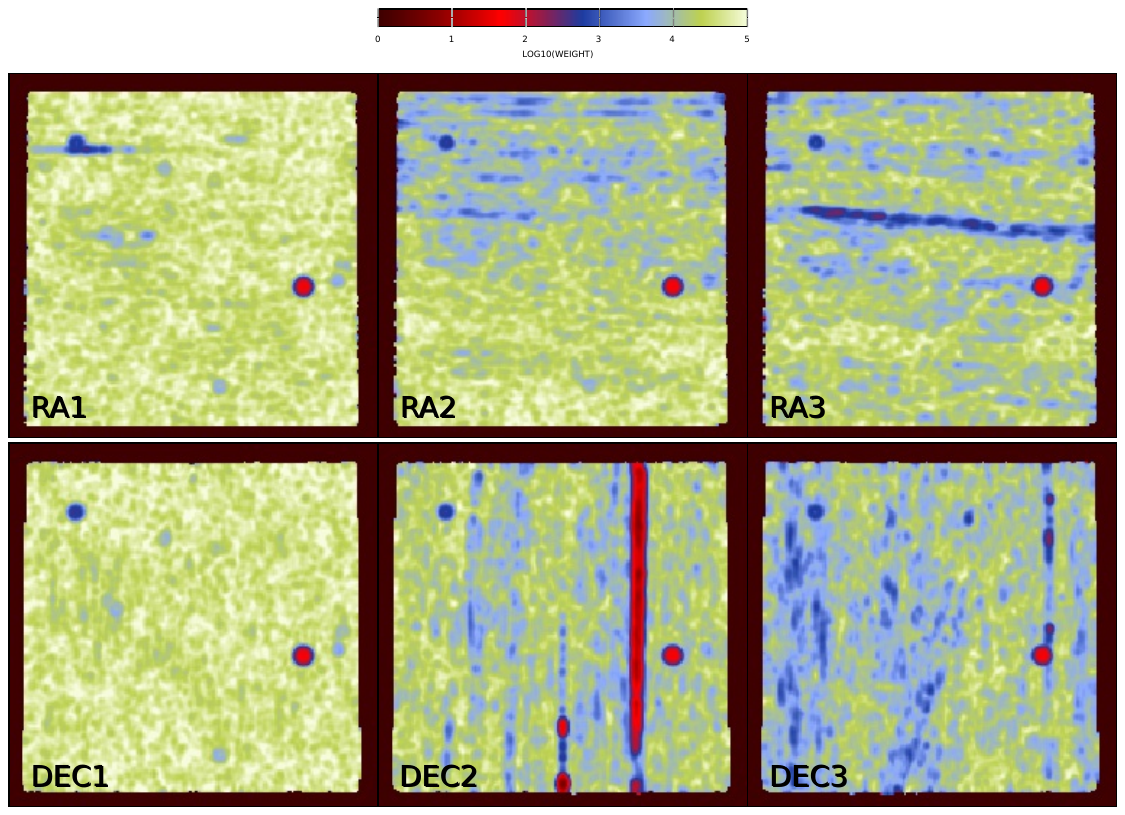}

    \caption{Images of the weights used for the stacking of the spectral cubes. The channel-0 image is smoothed with a Gaussian kernel of 4 pixels, and then weights are calculated as the inverse of the local variance of the brightness: $w_{(x,y)}=1/\sigma_{(x,y)}^2$. Note that isolated RFI and noisy regions are down weighted. Strong point sources are also down weighted but, since this occurs equally in all the images,  the weighted average is unaffected in these pixels.}
    \label{fig:weights}
\end{figure*}

\begin{figure*}
    \centering
    \includegraphics[width=0.75\textwidth,trim=25 425 25 15]{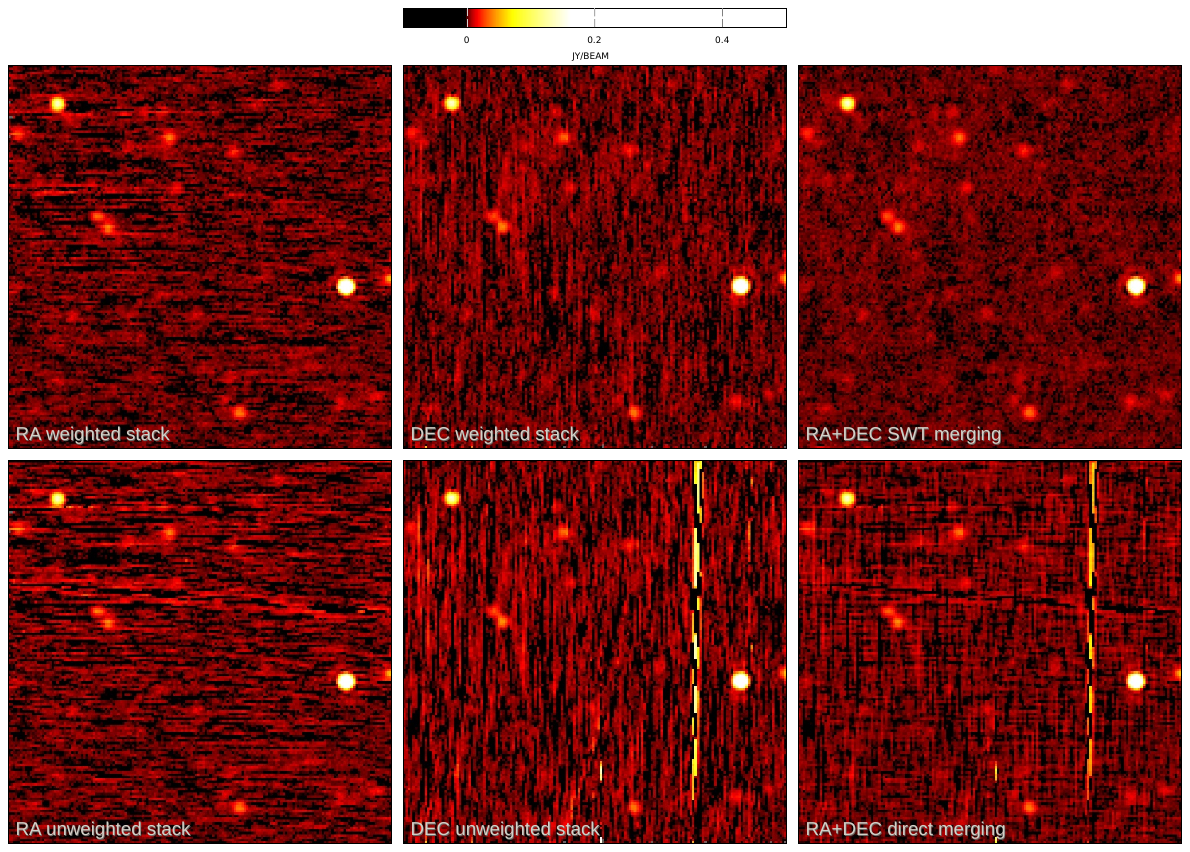}

    \caption{Top panels: weighted stacking of the RA (left) and DEC (middle) for polarization $R$ scans, shown in Fig.\ref{fig:radec}. The combination of the RA and DEC images obtained from the SWT merging is shown in the top-right panel. The unweighted stacking is shown in the bottom panels for comparison. The direct average of the unweighted RA and DEC scans shown in the bottom-right panel is evidently noisier than the SWT merging.}
    \label{fig:merging}
\end{figure*}

The SCUBE data reduction pipeline implements a weighted stacking imaging. For each polarization and Stokes parameter, spectral cubes are combined to increase the signal-to-noise ratio using a weighted average (Murgia \& Fatigoni 2023, ApJS, submitted). First, the channel-0 (spectral average) is smoothed with a Gaussian kernel of 4 pixels in FWHM and then the local intensity variance is calculated for each pixel over a patch of $3\times3$ FWHM in size. The weight at pixel $(x,y)$ is calculated as $w_{(x,y)}=1/\sigma_{(x,y)}^2$. In Fig.\,\ref{fig:radec} we show the spectral average of polarization R for six spectral cubes acquired on Feb-06-2016. The scans along RA1 and DEC1 are clearly better in terms of noise rms. Moreover, a strong RFI is present in the DEC2 scan and satellites streaks are visible in the RA3 and DEC3 scans. In Fig.\,\ref{fig:weights} we show the corresponding weights for these cubes calculated as described above. The isolated RFI and the noisy regions are down-weighted, as desired. Strong real sources, like Coma A, are down-weighted, but this does not represent a problem because the weights are nearly the same for all images and thus the average is consistently computed when dividing for the local sum of weights. The parameters used in the weighting procedure can be adjusted by the user, and also various weighting 
schemes can be selected. For instance, it is possible to weight both the channel-0 and the single-channels and then consider the minimum weight in order to suppress both broad-band and narrow-band RFI. In any case, the weights are always applied to all frequency channels in the spectral cubes before averaging them.
In the top-left and top-middle panels of Fig.\ref{fig:merging} we show the weighted stack of the three RA and DEC scans, respectively. The procedure is very effective in suppressing the RFI signals. The top-right panel show the merging of the orthogonal RA and DEC averages, obtained by mixing the Stationary Wavelet Transform detail coefficients \citep{murgia16}. This last step filters out the  correlated noise along the scan direction, and the final image is noticeably cleaner. For comparison, in the bottom panels we show the unweighted stack of the RA (bottom-left) and DEC (bottom-middle) scans along with their directed average (bottom-right). 
The noise of the R-pol image obtained with the weighted SWT merging is $\sigma_{R}=1.06$ mJy/beam (baselevel 0.03 mJy/beam) while the noise of the direct merging image is $\sigma_{R}=3.74$ mJy/beam (baselevel 0.46 mJy/beam).

\section{SRT C-band flux density consistency check.}
\label{appendixB}

\begin{figure*}
\centering
    \includegraphics[trim=50 150 50 150, width=0.45\textwidth]{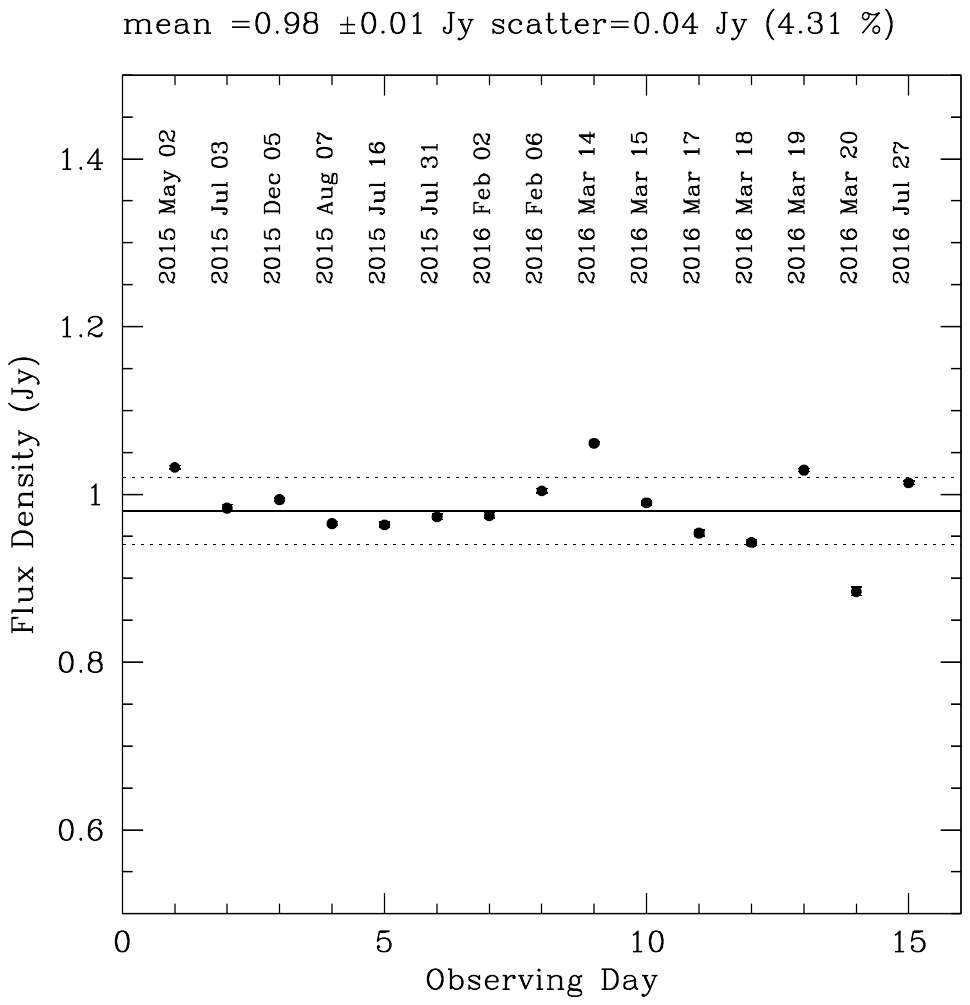}
\includegraphics[viewport=50 150 550 550,  width=0.45\textwidth]{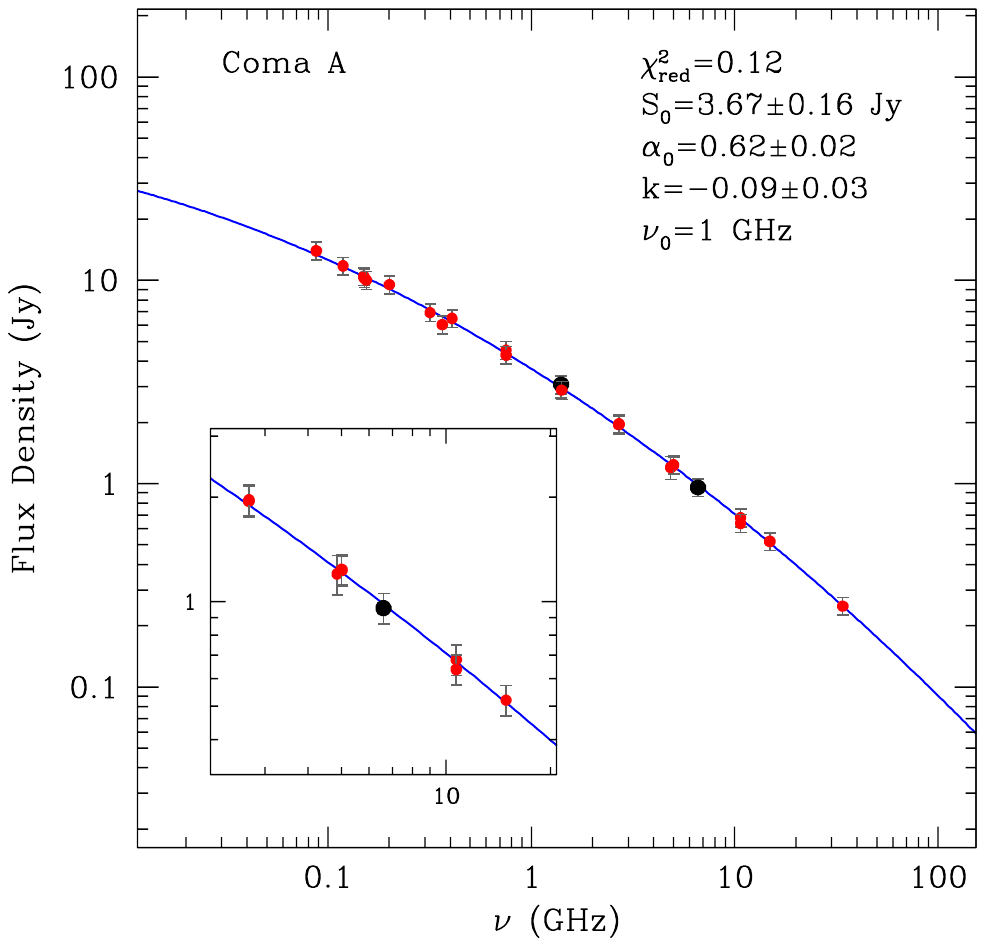}

    \caption{Left: Coma\,A Stokes I flux density at 6.6\,GHz measured for the different observing days. The scatter around the average (solid line) is of about 4\%, and it is represented by the dotted lines. Right: the SRT flux density measurements of Coma A at 1.4 and 6.6\,GHz (black dots) are compared with the literature data (red dots). The continuous blue line is the best fit of the modified power law model. The inset shows a zoom of the spectrum around 6.6\,GHz.}
    \label{fig:coma_a_check}
\end{figure*}

We evaluated the consistency of the SRT flux density scale at 6.6\,GHz by measuring the flux density of Coma A, that we assume to be constant in time during the period of our observations.
In the left panel of Fig.\,\ref{fig:coma_a_check} we show the flux density of Coma A at 6.6\,GHz measured during the campaign of observations. The mean value is of $0.98\pm 0.01$ Jy with a scatter of 0.04 Jy. The larger deviations, such as in the 20MAR16 session, are mostly due to higher RFI levels. The degree of reproducibility of the Coma A flux density can be considered as an estimate of the systematic error for a single-day SRT observation of about $\simeq 4\%$. However, by combining all the 15 observing days, the systematic error is expected to decrease by a factor of about $1/\sqrt{15}$ (assuming a Gaussian scatter).

In the right panel of Fig.\,\ref{fig:coma_a_check} we show the consistency of the Coma A flux density at 6.6\,GHz compared with the other measurement available in the literature (see Tab.\,\ref{tab:flux_comaa}). We modeled the global spectrum of Coma A with a modified power law (see Sect.\,\ref{sec:halo_spectrum}) characterized by three free parameters: the flux density at 1\,GHz, $S_0$, the spectral index at 1\,GHz, $\alpha$, and the spectral curvature, $k$. Using a reference frequency  $\nu_0=1$\,GHz, we found for the best-fit parameters: $S_0=3.67\pm0.16$\,Jy, $\alpha_0=0.62\pm0.02$, and $k=-0.09\pm0.03$.
The inset shows that the SRT data point is in remarkable agreement with both the measurements at nearby frequencies and the best fit of the modified power law model represented by the blue line. The difference between the model and the SRT measurement is indeed of about 2.5\%. This is somewhat larger than the systematic error of the flux density calibrator, which is about 1\% \citep{perley17}. Indeed, we assume a total systematic uncertainty of 3\% for the SRT observations at 6.6\,GHz. 

We also checked the SRT data point at 1.4\,GHz, $S_{\nu}=3.08 \pm0.05$ Jy. This data measurement is based on two observing days and at this frequency the difference between the model and the SRT measurement is indeed of about 11\%. 

\begin{table}
	\centering
	\caption{Flux density measurements of Coma\,A (3C\,277.3) taken from the literature using the NED. We obtained flux densities from the GLEAM \citep{wayth15, GLEAM} with AIPS task JMFIT.}
	\label{tab:flux_comaa}
	\begin{tabular}{ccl}
			\hline
          Frequency       & $S_{\nu}$              & Reference \\
		  (MHz)           & (Jy)                   &            \\
		\hline
		         87.5        & 13.96    $\pm$	 0.1		& GLEAM 72-103 \\
		         118.5	     & 11.78	$\pm$	 0.05		& GLEAM 103-134\\
		         150         & 10.4464  $\pm$ 	 1.0449		& TGSS CATS\\
		         151	     & 10.24    $\pm$	 0.4		& 7C \citep{waldram96}\\
		         154.5	     & 10.013	$\pm$	 0.04		& GLEAM 139-170 \\
		         200.5	     & 9.5144	$\pm$	 0.1		& GLEAM 170-231 \\
		         318	     & 6.94		$\pm$    0.31		& \citet{kuehr81}\\
		         365	     & 6.047	$\pm$	 0.133		& TEXAS \citep{douglas96}\\
		         408	     & 6.49		$\pm$    0.55		& \citet{kuehr81}\\
		         750	     & 4.3		$\pm$    0.20		& \citet{kuehr81}\\
		         750	     & 4.53		$\pm$    0.09		& \citet{kuehr81}\\
		         750	     & 4.28		$\pm$    0.08		& \citet{pauliny-toth66}\\
		         1400        & 3.08     $\pm$    0.05       & SRT this work\\
                  1400        & 3.06	 	$\pm$    0.11		& \citet{pauliny-toth66}\\
		         1400	     & 2.925 	$\pm$	 0.0978		& NVSS \citep{condon98}\\
		         1400	     & 2.933    $\pm$    0.082		& \citet{laing80}\\
		         1410	     & 2.88		$\pm$    0.03		& \citet{witzel79}\\
		         2695	     & 1.95		$\pm$    0.04		& \citet{kuehr81}\\
		         2695	     & 1.951    $\pm$	 0.04		& \citet{laing80}\\
		         2700	     & 1.97		$\pm$    0.1		& \citet{kuehr81}\\
		         4850	     & 1.202	$\pm$ 	 0.158		& 87GB \citep{gregory91}\\
		         5000	     & 1.23		$\pm$    0.06		& \citet{kuehr81}\\
		         5000	     &  1.24    $\pm$	 0.04		& \citet{laing80}\\
		         6600	     & 0.958	$\pm$	 0.001		& SRT this work\\
		         10700	     & 0.68		$\pm$    0.03 		& \citet{kellermann73}\\
		         10700	     & 0.638	$\pm$	 0.028		& \citet{laing80}\\
		         14900	     & 0.52	    $\pm$    0.03		& \citet{laing80}\\
		         34000	     & 0.250	$\pm$	 0.003		& \citet{lancaster05}\\
		        \hline
	\end{tabular}
\end{table}

\section{Halo polarized spot}
\label{appendixC}
\begin{figure*}
\centering
    \includegraphics[trim=20 550 0 0, width=1 \textwidth]{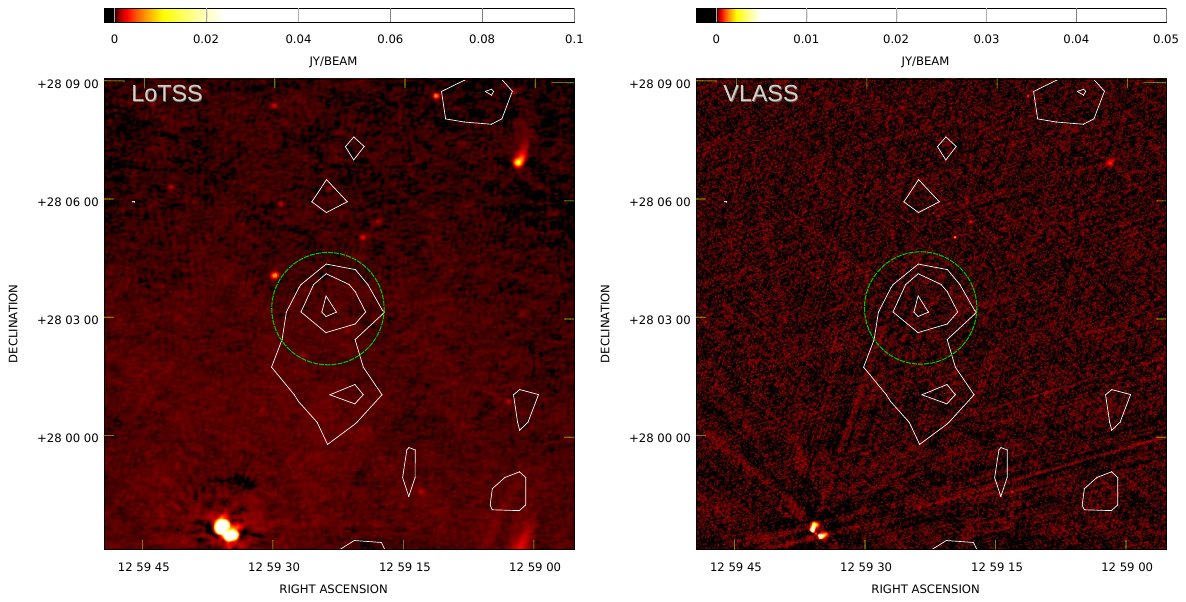}
    \caption{Coma radio halo. The SRT polarization contours at 6.6\,GHz are overimposed to the LoTSS image at 144\,MHz (left) and to the VLASS image at 2-4\,GHz (right). The green dash circle represents the FWHM beam of the SRT observation at 6.6\,GHz.}
    \label{fig:pol_spot}
\end{figure*}

We tested whether the polarized spot detected in the Coma radio halo could be due to a discrete radio source indistinguishable from the diffuse emission in the 6.6\,GHz SRT total intensity image. 
We inspected both the LoTSS at 144\,MHz and the high frequency VLASS images at 2-4\,GHz.The spot has a peak polarized intensity of 1.6\,mJy/beam at 6.6\,GHz. If the source is a 100\% polarized with an inverted spectral index $\alpha_{thick}=-2.5$ (the canonical value for a synchrotron self-absorption) then it would not be detectable in the total intensity at 144\,MHz, given the noise level of the LoTSS which is equal to 1$\sigma$ of 80 $\mu$Jy/beam. On the other hand, the source would still be detectable in total intensity in the VLASS at a significance of 3$\sigma$, since the 1$\sigma$ noise level in the VLASS mosaic is 70 $\mu$Jy/beam. However, there are no sources on the location of the polarized signal in either survey, see Fig.\,\ref{fig:pol_spot}. Also, the polarization level of this supposed source most likely should
be less than 100\%, resulting in a higher total intensity signal. We therefore exclude the possibility the polarized spot is caused by a contamination from a source with an inverted spectrum unresolved in total intensity by the SRT beam at 6.6\,GHz. 

\section{SRT beam area}
\label{appendixD}
\begin{figure*}
\centering
    \includegraphics[trim=20 250 50 0, width=0.45\textwidth]{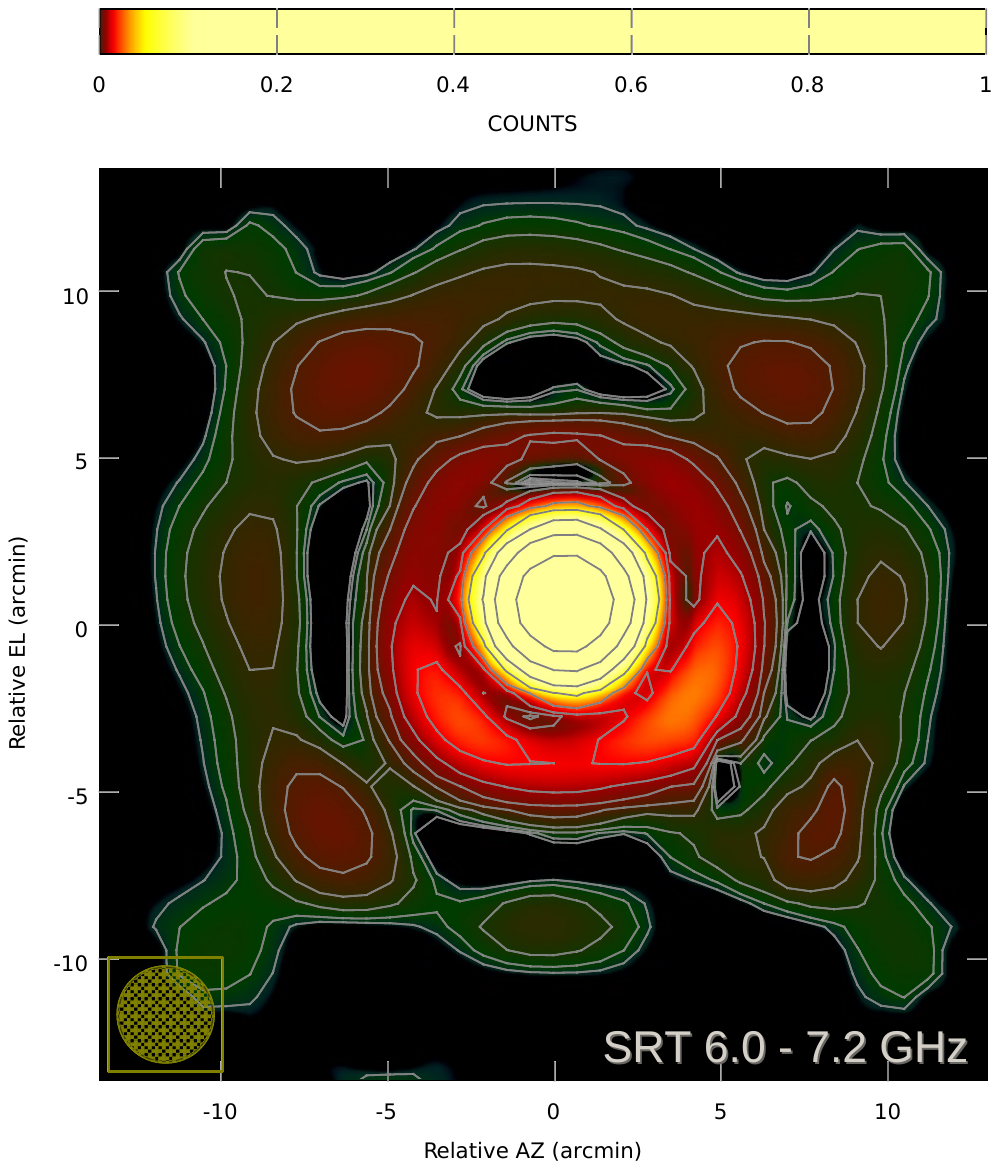}
    \includegraphics[viewport=50 170 550 550,  width=0.45\textwidth]{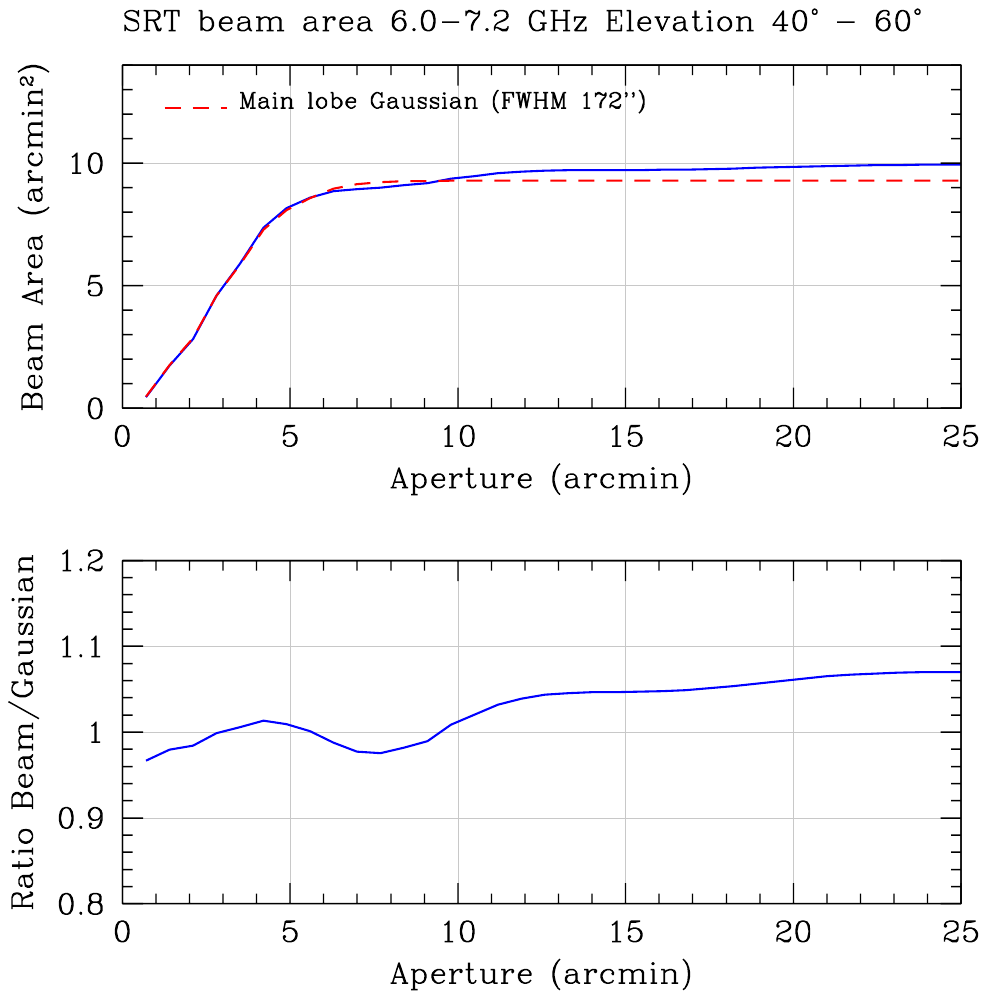}

    \caption{Left: SRT C-band beam in the band from 6.0 to 7.2\,GHz. Contours range from 0 to -40 dB and are spaced by -3 dB. Right: top panel shows the beam area as a function of the diameter of a circular aperture centered in the beam (continuous line). The dashed line is the area of a Gaussian fit to the main lobe of the beam. The bottom panel shows the ratio of the beam area to the Gaussian model as a function of the aperture diameter.}
    \label{fig:cband_beam}
\end{figure*}

The beam area is a relevant parameter for an accurate photometry of extended radio sources. In Fig.\,\ref{fig:cband_beam} we show the SRT beam in the 6.0 - 7.2\,GHz band, determined in the elevation range from $40^{\circ}$ to $60^{\circ}$ by observing strong point-like quasars. In the image are visible the main, the second (-20\,dB), and the third lobe (-30\,dB). The diagonal stripes are caused by the struts of the support structure of the secondary mirror of the telescope. 
A 2-dimensional Gaussian fit to the beam main lobe results in a FWHM of 172\,arcsec. In the top-right panel of Fig.\,\ref{fig:cband_beam} we present the profile of the beam area integrated over circular apertures of increasing diameter for both the observed beam (continuous line) and the Gaussian model (dotted line). The Gaussian model beam area tends asymptotically to $\Omega_{\rm beam}=2\pi/(8 \ln{2})\cdot {\rm FWHM}^2 \simeq 1.133\cdot{\rm FWHM}^2$, or 9.28\,arcmin$^2$. 
However, due to the contribution of the side-lobes the effective beam area is larger than the Gaussian beam area by about 7\% for apertures larger than 20\,arcmin, see bottom-right panel of Fig.\,\ref{fig:cband_beam}. The full beam area we consider for the aperture photometry is indeed  $\Omega_{\rm beam}=9.93$\,arcmin$^2$.

\bsp	
\label{lastpage}
\end{document}